  \documentclass{aa}
   \usepackage{graphics}

\begin{document}

   \voffset 3 true cm

   \thesaurus{10     
              (08.12.1; 08.19.1;  
               10.05.1;  
               10.19.1)}

   \title{Chemical enrichment and star formation in the Milky Way disk}
   \subtitle{II. Star formation history}

   \author{H. J. Rocha-Pinto\inst{1} 
           \and John Scalo\inst{2}
         \and W. J. Maciel\inst{1}
          \and Chris Flynn\inst{3}}

   \offprints{H. J. Rocha-Pinto}

   \institute{Instituto Astron\^omico e Geof\'{\i}sico, Universidade de S\~ao 
              Paulo, Av. Miguel Stefano 4200, 04301-904 S\~ao Paulo SP, 
              Brazil\\ emails: helio@iagusp.usp.br; maciel@iagusp.usp.br 
             \and Department of Astronomy, The University of Texas at Austin, 
              USA. email: parrot@astro.as.utexas.edu
             \and Tuorla Observatory, V\"ais\"al\"antie 20, FI-21500, Pikkii\"o, Finland. 
              email: cflynn@astro.utu.fi}
   
   \titlerunning{Chemical enrichment and formation of the Milky Way disk}
   \authorrunning{H.J. Rocha-Pinto et al.}

   \date{Received date; accepted date}
   
   \maketitle

   \begin{abstract}

   A chromospheric 
   age distribution of 552 late-type dwarfs 
   is transformed into a star formation history by the application of 
   scale height corrections, stellar evolutionary corrections and volume 
   corrections. We show 
   that the disk of our Galaxy has experienced enhanced episodes of star formation 
   at 0-1 Gyr, 2-5 Gyr and 7-9 Gyr ago, although the reality of the latter burst is 
   still uncertain. 
   The star sample birthsites are distributed over a very large range of
   distances because of orbital diffusion, and so give an estimate of the
   global star formation rate.  These results are compared with  
   the metal-enrichment rate, given by the age--metallicity relation, with the 
   expected epochs of close encounters between our Galaxy and the Magellanic Clouds, and 
   with previous determinations of the star formation history. Simulations are used to 
   examine the age-dependent smearing of the star formation history due to 
   age uncertainties, and the broadening of the recovered features, as well as to 
   measure the probability level that the history derived to be produced by statistical 
   fluctuations of a constant star formation history. We show, with a significance level greater 
   than 98\%, that the Milky Way have not had a constant star formation history. 

      \keywords{stars: late-type -- stars: statistics -- 
                Galaxy: evolution --
                solar neighbourhood}
   \end{abstract}

\section{Introduction}

   The question whether the Milky Way disk has experienced a smooth and constant 
   star formation history (hereafter SFH) or 
   a bursty one has been the subject of a number of studies since the initial 
   suggestions by Scalo (\cite{scalo87}) and Barry (\cite{barry}). Rocha-Pinto et al. 
   (\cite{RPSMC}; hereafter RPSMF) present a brief review about this question. There is evidence 
   for three extended periods of enhanced star formation in the disk. The use of the word 
   `burst' for these features (usually lasting 1-3 Gyr) is based on the fact that 
   all methods used to recover the SFH are likely 
   to smear out the original data 
   so that the star formation enhancement features could be narrower than they seem, 
   or be composed by a succession of 
   smaller bursts. In this sense, they were named bursts A, B and C, after Majewski 
   (\cite{majewski}). 

   The most efficient way to find the SFH is using the stellar 
   age distribution, which can be transformed into a star formation history after various 
   corrections. Twarog (\cite{twar}) summarized 
   some of these steps. Although his SFH is usually quoted as an evidence for the constancy 
   of the star formation in the disk, he states that during the most recent 4 Gyr, 
   the SFH has been more or less constant, followed by a sharp increase 
   from 4 to 8 Gyr ago, and a slow decline beyond that. His unsmoothed data were 
   also reanalysed by Noh \& Scalo (\cite{noh}) who have found more signs of 
   irregularity. 

   Barry 
   (\cite{barry}) 
   has improved this situation substantially by using chromospheric ages.  
   His conclusion was criticized by Soderblom et al. (\cite{soder91}), who 
   showed that the empirical data would be still consistent with a constant SFH if the 
   chromospheric emission--age relation is suitably modified. However, 
   Rocha-Pinto \& Maciel (\cite{RPM98}) have recently argued that the scatter 
   in Soderblom et al. (\cite{soder91})'s Figure 13, which is the main 
   feature that could suggest a non-monotonic age calibration, is probably caused 
   by contamination in the photometric indices due to the chromospheric 
   activity. The chromospheric activity--age relation 
   was also further investigated by Donahue (\cite{don93}, \cite{don98}), and 
   the new proposed calibration still predicts a non-constant SFH if applied 
   to Barry's data.

   The SFH derived in this paper is based on a new chromospheric sample 
   compiled by us 
   (Rocha-Pinto et al. \cite{paper1}, hereafter Paper I). 
   This paper is organized as follows: In section 2, we address the transformation of 
   the age distribution into SFH. The results are presented in section 3. In section 4, 
   statistical 
   significances for the SFH are provided by means of a number of simulations. 
   The impact of the age errors on the recovered SFH is also studied. Some comparisions with 
   observational constraints are addressed in section 5, and each particular feature 
   of the SFH is discussed in section 6, in view of the results from the simulations and 
   comparisons with other data. The case for a non-monotonic chromospheric activity--age relation 
   is discussed in section 7. Our final 
   conclusions follow in section 8. A summary of this work was presented in RPSMF.

\section{Converting age distribution into SFH}

     Assuming that the sample under study is representative of the galactic disk, the 
     star formation rate can be derived from its age distribution, since 
     the number of stars in each age bin is supposed to be correlated with the number 
     of stars initially born at that time. 

     We use the 
     same 552 stars with which we have derived the AMR (Paper I), after correcting 
     the metallicities 
     of the active stars for the $m_1$ deficiency (Gim\'enez et al. \cite{gimenez}; 
     Rocha-Pinto \& Maciel \cite{RPM98}), which accounts for the influence of the 
     chromospheric activity on the photometric indices. The reader is referred to Paper I for details 
     concerning the sample construction and the derivation of ages, from the chromospheric 
     Ca H and K emission measurements. 

     The transformation of the chromospheric 
     age distribution into history of the 
     star formation rate comprises three intermediate corrections, 
     namely the volume, evolutionary and scale height corrections. They are explained 
     in what follows.

  \subsection{Volume correction}

     Since our sample is not volume-limited, there could be a bias in the relative number 
     of stars in each age bin: stars with different chemical compositions have different 
     magnitudes, thus the volume of space sampled varies from star to star. To correct 
     for this effect, before counting the number of stars in each age bin, we have 
     weighted each star (counting initially as 1) by the same factor $d^{-3}$ used 
     for the case of the AMR, where $d$ is the maximum distance at which the star would 
     still have apparent magnitude lower than a limit of about 8.3 mag (see Paper I for details).
     
      \begin{figure}
      \resizebox{\hsize}{!}{\includegraphics{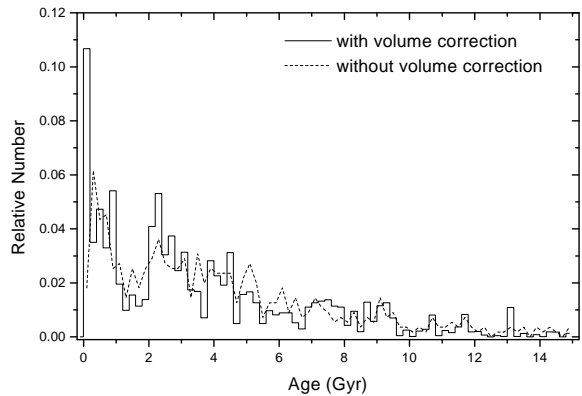}}
      \caption[]{Chromospheric age distribution with and without volume correction, which 
         was applied to our sample to allow the derivation of a magnitude-limited SFH.}
      \label{volume}
      \end{figure}

     This correction proves to change significantly the age distribution as can be seen 
     in Figure \ref{volume}. 

  \subsection{Evolutionary corrections}
     
\begin{figure}
      \resizebox{\hsize}{!}{\includegraphics{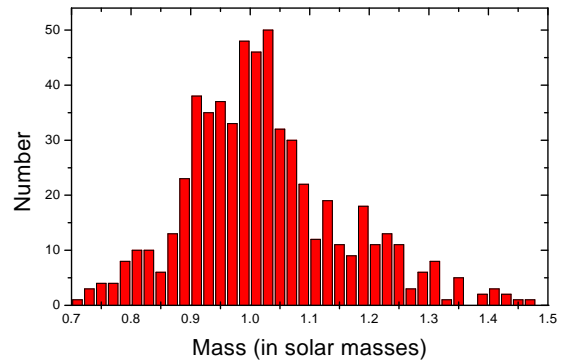}}
      \caption[]{Mass distribution of the sample. Masses were calculated from a mean 
          mass--magnitude relation given by Scalo (\cite{scalo}). From the figure, we 
          estimate a mass range of 0.8-1.4 $M_\odot$ for our sample. Note the substantial 
          absence of massive stars, compared to the left wing of the mass distribution. 
          The evolutionary corrections attempt to alleviate this bias.}
      \label{mass}
      \end{figure}

     A correction due to stellar evolution is needed when a sample comprises stars 
     with different masses. The more massive stars have a life expectancy lower than 
     the disk age, thus they would be missing in the older age bins. The mass of our 
     stars was calculated from a characteristic mass--magnitude relation for the solar 
     neighbourhood (Scalo \cite{scalo}). In Figure \ref{mass}, the mass distribution is 
     shown. We take the mass range of our sample as 0.8 to 1.4 $M_\odot$, which 
     agrees well with the spectral-type range of the sample from nearly F8 V to K1-K2 V. 
     As an example for the necessity of these corrections, the stellar lifetime of a 
     1.2 $M_\odot$ is around 5.5 Gyr (see Figure \ref{stellife} below). This means 
     that only the most recent age bins are expected to have stars at the whole mass 
     range of the sample.

      \begin{figure*}
      \resizebox{\hsize}{!}{\includegraphics{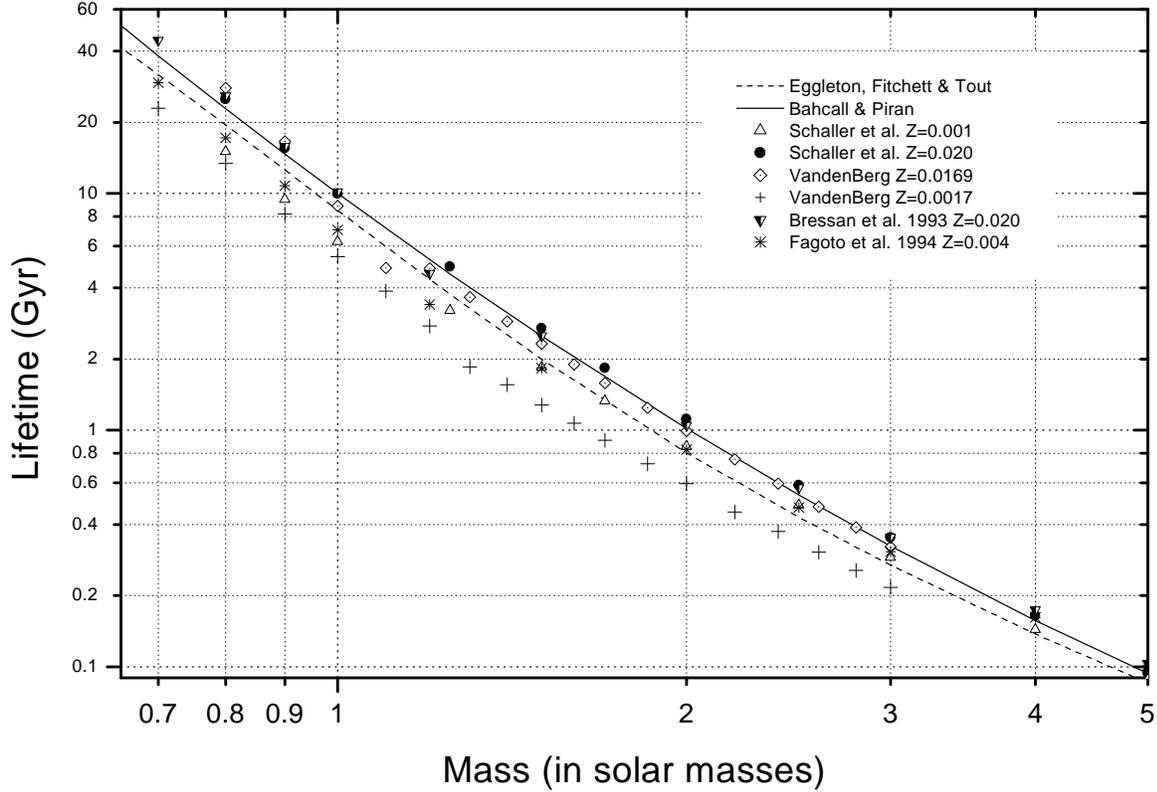}}
      \caption[]{Stellar Lifetimes from a variety of sources: Bahcall \& Piran 
          (\cite{bahcall}); VandenBerg (\cite{VandenBerg}), for $Z=0.0169$ and $Z=0.0017$; 
          Eggleton, Fitchett \& Tout (\cite{EFT}); Schaller et al. (\cite{schallera}), for 
          $Z=0.02$ and $Z=0.001$; Bressan et al. (\cite{bressan}), for $Z=0.02$; 
          Fagotto et al. (\cite{fagotto}a), for $Z=0.004$.}
      \label{stellife}
      \end{figure*}

     The corrections are given by the following formalism. The number of stars born at 
     time $t$ ago (present time corresponds to $t=0$), with mass 
     between 0.8 and 1.4 $M_\odot$ is
    \begin{equation}
      N^\ast (t)=\psi(t)\int^{1.4}_{0.8} \phi(m)\, dm,
    \label{Nast}
    \end{equation}
     where $\phi(m)$ is the initial mass function, assumed constant, 
     and $\psi(t)$ is the star formation rate 
     in units of $M_\odot$~Gyr$^{-1}$pc$^{-2}$. The number of these objects that have already died 
     today is 
    \begin{equation}
      N^\dag (t)=\psi(t)\int^{1.4}_{{m_\tau(t)}} \phi(m)\, dm,
    \label{Ndied}
    \end{equation}
    where $m_\tau(t)$ is the mass whose lifetime corresponds to $t$. From these equations, 
    we can write that the number of still living stars, born at time $t$, as 
    \begin{equation}
      N^{\rm obs} (t)= N^\ast (t)- N^\dag (t).
    \label{Nobs}
    \end{equation}
    Using equations (\ref{Nast}) and (\ref{Ndied}), we have
    \begin{equation}
      N^\dag (t)= \left[{\int^{1.4}_{{m_\tau(t)}} \phi(m)\, dm \over \int^{1.4}_{0.8} 
         \phi(m)\, dm}\right] N^\ast (t) = {\alpha(t)\over\beta}N^\ast (t);
    \label{fractions}
    \end{equation}

    \begin{equation}
       N^\ast (t) = \varepsilon(t) N^{\rm obs} (t),
     \label{link}
    \end{equation}
    where
    \begin{equation}
        \varepsilon(t)=\left(1-{\alpha(t)\over\beta}\right)^{-1}.
     \label{equeta}
    \end{equation}

    The number of objects initially born at each age bin can be calculated by using 
    equation (\ref{equeta}), so that we have to multiply the number of stars presently 
    observed by the $\varepsilon$ factor. These corrections were independently developed 
    by Tinsley (\cite{tinsley74}), in a different formalism. RPSMF 
    present another way to express this correction in terms of the 
    stellar lifetime probability function. We stress that all these formalisms 
    yield identical results.

    The function $m_\tau(t)$ can be calculated by inverting stellar lifetimes 
    relations. Figure \ref{stellife} shows stellar lifetimes for a number of 
    studies published in the literature. Note the good agreement between the relations 
    of the Padova group (Bressan et al. \cite{bressan}; Fagotto et al. \cite{fagotto}a,b) 
    and that by Schaller et al. (\cite{schallera}), as well with Bahcall \& 
    Piran (\cite{bahcall})'s lifetimes. The stellar lifetimes for $Z=0.0017$ given 
    by VandenBerg (\cite{VandenBerg}) are underestimated probably due to the 
    old opacity tables used by him. The agreement in the stellar lifetimes shows 
    that the error introduced in the SFH due to the evolutionary corrections is not 
    very large.

    The adopted turnoff-mass relation was calculated from the stellar lifetimes by 
    Bressan et al. (\cite{bressan}) and Schaller et al. (\cite{schallera}), for solar 
    metallicity stars:
    \begin{equation}
      \log m_\tau(t) = 7.59 - 1.25\log t + 0.05(\log t)^2,
     \label{mtbressan}
    \end{equation}
    where $t$ is in yr. This equation is only valid for the mass 
    range $5 M_\odot > m > 0.7 M_\odot$.
 
    We have also considered the effects of the metallicity-dependent lifetimes on 
    the turnoff mass. To account for this dependence, we have adopted the 
    stellar lifetimes for different chemical compositions, as given by 
    Bressan et al. (\cite{bressan}) and Fagotto et al. (\cite{fagotto}a,b). 
    Equations similar to Eq. (\ref{mtbressan}) 
    were derived for each set of isochrones and the metallicity dependence of the 
    coefficients was calculated. We arrive at the following equation:
    \begin{equation}
     \log m_\tau(t) = a + b\log t + c(\log t)^2,
       \label{mtzdep}
    \end{equation}
    where $a = 7.62  - 1.56{\rm [Fe/H]}$, $b = -1.26+0.34{\rm [Fe/H]}$, 
    $c=0.05-0.02{\rm [Fe/H]}$. Since [Fe/H] depends on time we use a 
    third-order polynomial fitted to the AMR derived in Paper I. In that work, we 
    have also shown that the AMR is very affected at older ages, due to the 
    errors in the chromospheric bins. The real AMR must be probably steeper, 
    and the disk initial metallicity around $-0.70$ dex. The effect of this in the 
    SFH is small. The use of a steeper AMR increases the turnoff mass at older 
    ages, decreasing the stellar evolutionary correction factors (Equation 
    \ref{equeta}). As a result, the SFH features at young and intermediate age bins 
    (ages lower than 8 Gyr) increases slightly related to the older features, in units of 
    relative birthrate which is the kind of plot we will work in the next sections. 
        
    Note that equation (\ref{mtzdep}) does not 
    reduce to equation (\ref{mtbressan}) when $\rm{[Fe/H]} = 0$. The former  
    was calculated from an average between two solar-metallicity stellar evolutionary 
    models, while the latter uses the results of the same model with varying 
    composition. The difference in the turnoff mass from these equations 
    amount 12-15\% from 0.4 to 15 Gyr.

    The initial mass function (IMF) also enters in the formalism of the $\varepsilon$ 
    factor. For the mass range under consideration, the IMF depends on the SFH, more 
    specifically on the present star formation rate. It could be derived from 
    open clusters, but they are probably severely affected by mass segregation, 
    unresolved binaries and so on (Scalo \cite{scalo98}). We have adopted
    the IMF by Miller \& Scalo (\cite{ms79}), for a constant SFH, which gives an average 
    value for the mass range under study. Power-law IMFs 
    were also used to see the effect on the results. 

\begin{figure}
      \resizebox{\hsize}{!}{\includegraphics{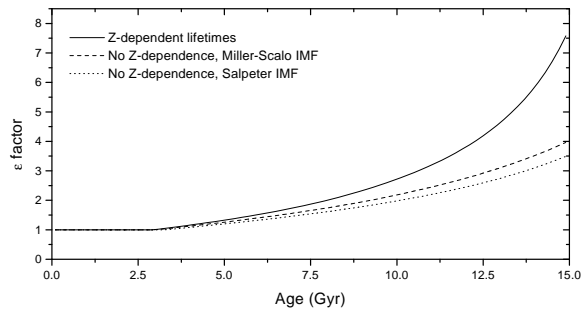}}
      \caption[]{Stellar evolution correction factors. The curves stand 
    for Equations (\ref{mtbressan}; dashed line) and (\ref{mtzdep}; solid line) 
    and Miller-Scalo's IMF. 
    A third curve (dotted line) gives the results of using a Salpeter IMF with the turnoff-mass 
    given by Equation (\ref{mtbressan}).}
      \label{epsilon}
      \end{figure}

    In Figure \ref{epsilon} we show how this factor varies with age. The curves 
    represent Equations (\ref{mtbressan}; dashed curve) and 
    (\ref{mtzdep}; solid curve) using the Miller-Scalo's IMF. 
    A third curve (shown by dots) gives the results using a Salpeter IMF with the turnoff-mass 
    given by Equation (\ref{mtbressan}). The $\varepsilon$ factor does not vary 
    very much when we use a different IMF. Being flatter than Salpeter IMF, the 
    correction factors given by the Miller-Scalo IMF are higher. However, the effects 
    of neglecting the metallicity-dependence of the stellar lifetimes are much more 
    important in the calculation of this correcting factor. Since low-metallicity 
    stars live less than their richer counterparts, the turnoff-masses at older ages 
    are highly affected. In the following section, we will use the $\varepsilon$ 
    factors calculated for metallicity-dependent lifetimes.

  \subsection{Scale height correction}

    Another depopulation mechanism, affecting samples limited to the galactic plane, is 
    the heating of the stellar orbits which increases the scale heights of the older 
    objects. To correct for this we use the following equations. Assuming that the 
    scale heights in the disk are exponential, the transformation of the observed 
    age distribution, $N_0(t)$, into the function $N(t)$ giving the total number of 
    stars born at time $t$ is
    \begin{equation}
      N(t)=2H(t)N_0(t),
      \label{scale}
    \end{equation}
    where $H(t)$ is the average scale height as a function of the stellar age. A problem 
    arises since scale heights are always given as a function of absolute magnitude 
    or mass. To solve for this, we use an average stellar age corresponding to a given 
    mass, following the iterative procedure outlined in Noh \& Scalo (\cite{noh}). 
    This average age, $\langle\tau\rangle$, can be obtained by
    \begin{equation}
       \langle\tau\rangle={\int^{{\tau_m}}_0 t N(t)\,dt\over \int^{{\tau_m}}_0 
           N(t)\,dt};
     \label{taumedio}
    \end{equation}
    where $\tau_m$ is the lifetime of stars having mass $m$, and $N(t)$ is the star 
    formation rate. Since $\langle\tau\rangle$ depends on the star formation rate, 
    which on the other hand depends on the average ages through the definition of 
    $H(t)$, equations (\ref{scale}) and (\ref{taumedio}) can only be solved by iteration. 
    We use the chromospheric age distribution as the first guess $N_0(t)$, and calculate 
    the average ages $\langle\tau\rangle_0$. These are used to convert $H(m)$ to $H(t)$, 
    and the star formation history is found by equation (\ref{scale}), giving $N_1(t)$. 
    This quantity is used to calculate $\langle\tau\rangle_1$ and a new star formation 
    rate, $N_2(t)$. Note that, in equation (\ref{scale}), the quantity that varies in 
    each iteration is $H(t)$, not the chromospheric age distribution $N_0(t)$. Our 
    calculations have shown that convergence is attained rapidly, generally after the 
    second iteration.

\begin{figure}
      \resizebox{\hsize}{!}{\includegraphics{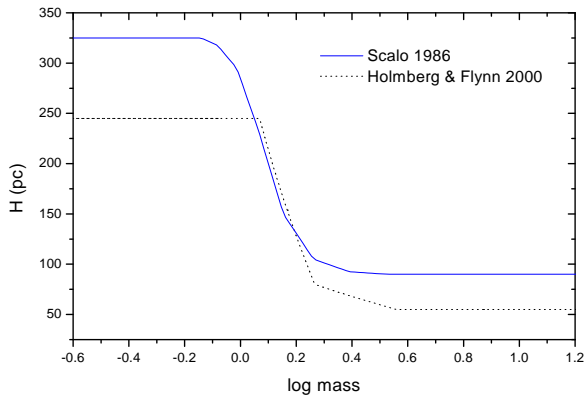}}
      \caption[]{Scale heights given by Scalo (\cite{scalo}, solid line) and 
      Holmberg \& Flynn (\cite{holmberg}, dotted line).}
      \label{scaleheight}
      \end{figure}

    Great uncertainties are still present in the scale heights for disk stars. Few  
    works have addressed them since Scalo (\cite{scalo})'s review (see e.g., Haywood, 
    Robin \& Crez\'e \cite{hay}). We will be working with two different scale heights: 
    Scalo (\cite{scalo}) and Holmberg \& Flynn (\cite{holmberg}), that are shown in Figure 
    \ref{scaleheight}. Haywood et al.'s scale heights are just in the middle of these, so 
    they set the limits on the effects in the derivation of the SFH.

    The major effect of the scale heights is to increase the contribution of the older 
    stars in the SFH. Better scale heights would not change significantly the results, so 
    that we limit our discussion to these two derivations.

  \section{Star formation history in the galactic disk}

  \subsection{Previous chromospheric SFH determinations}

      \begin{figure}
      \resizebox{\hsize}{!}{\includegraphics{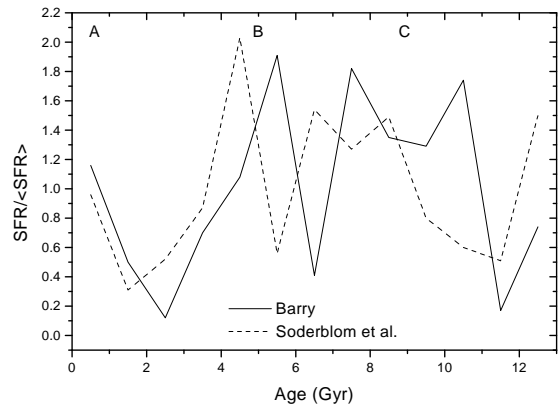}}
      \caption[]{Comparison between chromospheric SFHs published in the literature: Barry 
         (\cite{barry}, according to Noh \& Scalo \cite{noh}) and Soderblom et al. 
         (\cite{soder91}, according to Rana \& Basu \cite{ranabasu}). 
         The position of bursts A, B and C (named after Majewski \cite{majewski}) 
           are marked.}
      \label{SFRs}
      \end{figure}

     In Figure \ref{SFRs}, we show a comparison between two SFHs, derived 
     from chromospheric age distributions available in the literature: 
     Barry (\cite{barry}, SFH given by Noh \& Scalo {\cite{noh}) and Soderblom et al. 
     (\cite{soder91}, SFH given by Rana \& Basu \cite{ranabasu}). In this plot, as well 
     as in subsequent figures, the SFH will be expressed always as a relative birthrate, 
     which is defined as the star formation rate in units of average past star formation 
     rate (see Miller \& Scalo \cite{ms79}, for rigorous definition). 

     Note that the SFHs in Figure \ref{SFRs} are very 
     similar to each other, a result not really surprising since Soderblom et al. 
     have used the same sample used by Barry. On the other hand, the corresponding  
     events in Barry's SFH appears 1 Gyr earlier in Soderblom et al.'s SFH. 
     The different age calibrations 
     used in these works are the sole cause of this discrepancy. Barry makes 
     use of Barry et al. (\cite{barry87})'s calibration which used a low-resolution index 
     analogous to Mount Wilson $\log R'_{\rm HK}$, while Soderblom et al. use a  
     calibration derived by themselves. In Figure \ref{calib}, we 
     show a comparison of the ages for Barry 
     (\cite{barry})'s stars using both age calibrations. The difference in the ages are     
     clearly caused by the slopes of the calibrations. Barry et al. (\cite{barry87})'s 
     calibration gives higher ages compared to the other calibration, which explains 
     the differences in the corresponding SFHs published.

      \begin{figure}
      \resizebox{\hsize}{!}{\includegraphics{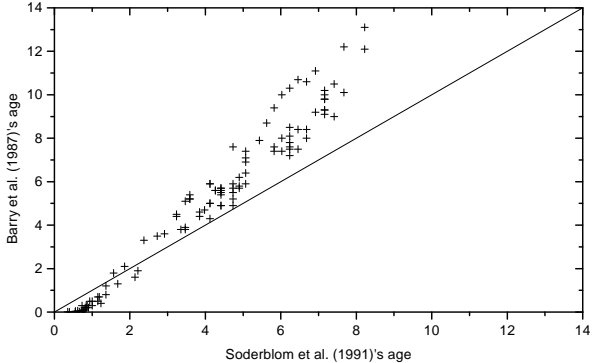}}
      \caption[]{Comparison of stellar ages (in Gyr) in the 
             calibrations by Barry et al. (\cite{barry87})'s and 
         Soderblom et al. (\cite{soder91}). The first age calibration seems to 
         overestimate the chromospheric ages by around 1 Gyr.}
      \label{calib}
      \end{figure}

  \subsection{Determination of the SFH}

    The three corrections described in section 2 are applied to our data in 
    the following order: the age distribution 
    is first weighted according to the volume corrections, then each age bin is 
    multiplied by the $\varepsilon$ factor and we iterate the result according to 
    equations 
    (\ref{scale}) and (\ref{taumedio}). The final result is the best estimate of 
    the star formation history. It is shown in Figure 
    \ref{sfrh}a, for an age bin of 0.4 Gyr and Scalo's scale height. There can be seen 
    three regions where the stars are more concentrated: at 0-1 Gyr, 
    2-5 Gyr and 7-9 Gyr ago. 
    Beyond 10 Gyr of age, the SFH is very irregular, probably reflecting more the 
    sample incompleteness in this age range, and age errors, than real features. 
    These patterns are 
    still present even considering a smaller age bin of 0.2 Gyr. Figure \ref{sfrh}b 
    shows the same for Holmberg \& Flynn (\cite{holmberg}) scale heights. The only 
    difference comes from the amplitude of the events. In this plot, the importance 
    of the older bursts is increased, since in Holmberg \& Flynn (\cite{holmberg}) 
    the difference in the scale heights of the oldest to the youngest stars is 
    greater than the corresponding value in Scalo's scale heights.

      \begin{figure}
      \resizebox{\hsize}{!}{\includegraphics{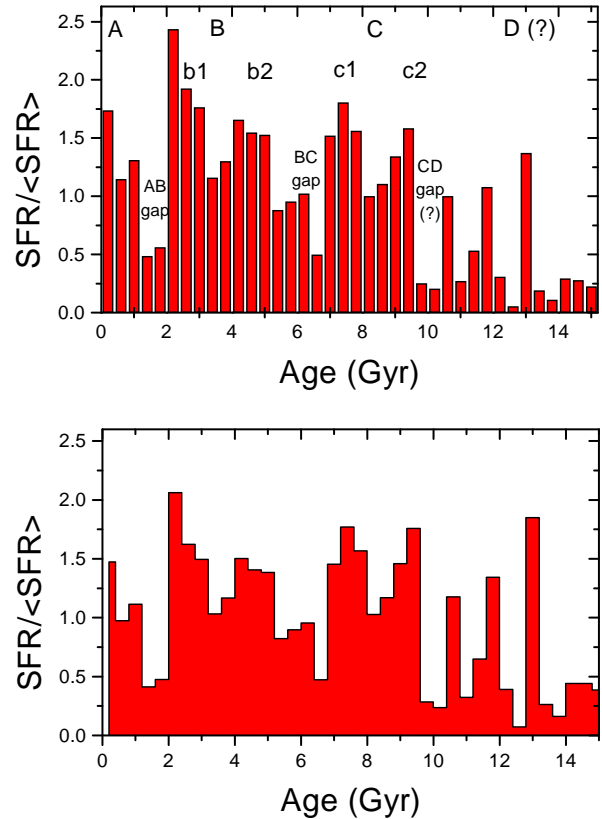}}
      \caption[]{Star formation rate for an age bin of 0.4 Gyr. The nomenclature 
      used by Majewski (\cite{majewski}) was extended to be used with the main features 
      of the SFH. The terms B1 and B2, and C1 and C2, stand for substructures of the 
      supposed bursts B and C, respectively. Also shown is the supposed burst D. The 
      gaps between the peaks are named AB gap, BC gap, and so on. The upper and 
      lower panels show the SFH using Scalo (\cite{scalo}) and Holmberg \& 
      Flynn (\cite{holmberg}) scale heights, respectively.}
      \label{sfrh}
      \end{figure}

    We have used an extended nomenclature to that of Majewsky (\cite{majewski}) 
    to refer to the features found. At the age range where bursts B and C were 
    thought to occur double-peaked structures are now seen. 
    Thus, we have used the terms B1 and B2, and C1 and C2, to these 
    substructures. Also shown is the supposed burst D, as Majewski (\cite{majewski}) 
    had suggested. Their meaning will be discussed later. The lulls 
    between the bursts were named AB gap, BC gap and so on. Some of us have 
    previously referred to the most recent lull as `Vaughan-Preston gap'. We now 
    avoid the use of this term because:
    \begin{enumerate}
    \item The Vaughan-Preston gap is a feature in the {\it chromospheric 
       activity distribution};
    \item Due to the metallicity-dependence of the age calibration, the 
       Vaughan-Preston gap is not linearly reflected in an age gap;
    \item Henry et al. (\cite{HSDB}, hereafter HSDB) shows that the Vaughan-Preston gap 
       is less pronounced than was earlier thought, and does not resemble a gap 
       but a transition zone.
    \end{enumerate}

    Comparing with other studies in the literature, the SFH seems particularly 
    different. There are still three major star formation episodes but 
    their amplitude, extension and time of 
    occurrence are not identical to those that were previously found by other 
    authors. Table \ref{simi} summarizes the main characteristics of our SFH 
    comparing to that of Barry (\cite{barry}, as derived 
    in Noh \& Scalo \cite{noh}). In the Table, the entries with two values stand 
    for the SFH derived with different scale heights. The first number refers to 
    the SFH with Scalo's scale height, and the other refers to that with Holmberg 
    \& Flynn's.

     \begin{table*}
      \caption[]{Main features of the SFH compared with Barry (\cite{barry}).}
         \label{simi}
         \begin{flushleft}
    {\halign{%
    #\hfil & \qquad\qquad\hfil#\hfil & \qquad\qquad\hfil#\hfil \cr
    \noalign{\hrule\medskip}
    & This work & Barry (\cite{barry}) \cr
    \noalign{\medskip\hrule\medskip}
         Number of `bursts' & 3 & 3 \cr
         Age of burst A & 0-1 Gyr & 0-1 Gyr \cr
         Age of burst B & 2-5 Gyr & 4-6 Gyr \cr
         Age of burst C & 7-9.5 Gyr & 7-11 Gyr \cr
         Stronger burst & B & B \cr
         Duration of the most recent lull (AB gap) & 1 Gyr & $\le$ 3 Gyr \cr
         (\% of stars formed in burst A)/Gyr & 10.48/9.58 & 8.92 \cr
         (\% of stars formed in burst B)/Gyr & 10.40/9.72 & 11.50 \cr
         (\% of stars formed in burst C)/Gyr & 8.88/10.88 & 11.92 \cr
  \noalign{\medskip\hrule}}}
         \end{flushleft}
   \end{table*}

    As we can see, the main events of our SFH seem to occur earlier than the 
    corresponding events in Barry's SFH, by approximately 1 Gyr. This can 
    also be seen in Figure \ref{SFRs}: the SFR from Soderblom et al. 
    (\cite{soder91})'s 
    data have features earlier than Barry by about 1 Gyr. This comes mainly 
    from the use of Soderblom et al. (\cite{soder91})'s age calibration on which we have based 
    our ages. This hypothesis is reinforced by the fact that the fraction of the 
    stars formed in each burst is in reasonable agreement with the corresponding events 
    in Barry's SFH (see Table \ref{simi}). The events we have found are most likely 
    to be the same that have appeared in previous works, and the difference in the time 
    of occurrence comes from 
    the shrinking of the chronologic scale of the age calibration.

    The narrowing of the AB gap is one of the main differences of our SFH and 
    that found by Barry. This can be expected 
    since our sample does not show a well-marked Vaughan-Preston gap, contrary to what is 
    found in the survey of 
    Soderblom (\cite{soder}), from which Barry (\cite{barry}) selected his sample.

    Some other differences in the amplitude and duration of the bursts can be 
    understood as resulting from the differences in the samples used 
    by us and by Barry. Nearly 70\% of our stars come from HSDB survey. We 
    have already shown in Paper I that HSDB and Soderblom (\cite{soder}) surveys have 
    different chromospheric activity distributions. These are directly reflected in 
    the SFH. 

    We have found double peaks at bursts B and C. At the present moment we cannot 
    distinguish these features from a real double-peaked burst (that is, two unresolved 
    bursts) or a single smeared 
    peak. However, it is interesting to see that the previous chromospheric SFHs give 
    some evidence for a double burst C. In Figure \ref{SFRs} burst C also seems 
    to be formed by two peaks. 
    On the other hand, the same does not occur for burst B. The feature called B2 
    corresponds more closely to burst B in the previous studies, but at the age 
    where we have found B1, the other SFHs show a gap.

    The resulting SFH comes directly from the age distribution, in an approach which assumes 
    that the most frequent ages of the stars indicate the epochs when the star formation 
    was 
    more intense. Both the evolutionary and the scale height corrections do not 
    change the clumps of stars already present in the age distribution. The only correction 
    which could introduce spurious patterns in it is the volume correction, which must 
    be applied before the other two. Figure \ref{volume} shows how it affects the age 
    distribution. It is elucidating that the major patterns of the age distributions are not 
    much changed after this correction. We refer basically to the clumps of stars 
    younger than 1 Gyr and stars with ages between 2 and 4 Gyr. These clumps will be         
    identified with burst A and B, respectively, after the application of the other 
    corrections. Note also, that the AB gap is clearly seen in the age distribution 
    before the volume correction. In spite of it, it is necessary to know if the presence 
    of stars with very high weights (due to their proximity and low temperature) could 
    affect the results. Therefore, we have recalculated the SFH now disregarding the stars 
    that have very high weights after the volume correction. We have cut the sample to 
    those stars with weights not exceeding 2$\sigma$ and 3$\sigma$. The resulting SFHs is 
    compared to the SFH of the whole sample in Figure \ref{outliers}. It is possible to 
    see that the presence of outliers does not affect the global result. The 
    uncertainty introduced affects mainly the amplitude of the events, at a level similar to 
    that introduced by the uncertainty in the scale heights. We believe that 
    the volume correction has not impinged artificial patterns on the data, and 
    that the star formation just derived reflects directly the observed distribution 
    of stellar ages in the solar vicinity.

      \begin{figure}
      \resizebox{\hsize}{!}{\includegraphics{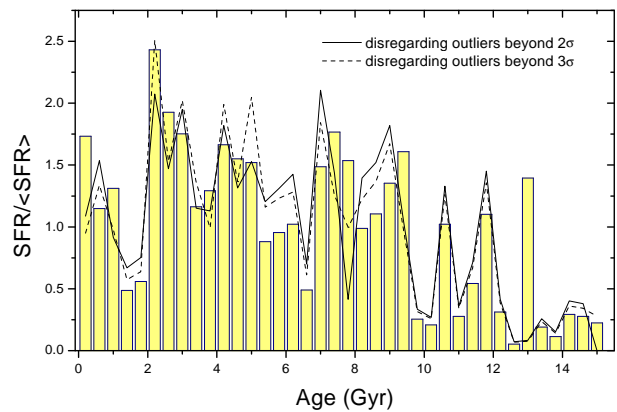}}
      \caption[]{Star formation rate calculated after disregarding outliers with 
      weights exceeding 2$\sigma$ (solid line) and 3$\sigma$ (dashed line), after 
      the volume corrections, compared to 
      the history used throughout this paper.}
      \label{outliers}
      \end{figure}

  \section{Statistical significance of the results}

  \subsection{Inconsistency of the data with a constant SFH}

    There is a widespread myth on galactic evolutionary studies about the near constancy 
    of the SFH in the disk. This comes primarily from earlier studies setting constraints to 
    the present relative birthrate (e.g., Miller \& Scalo \cite{ms79}; Scalo 
    \cite{scalo}).  
    The observational constraints have favoured a value near unity, 
    and that was interpreted 
    as a constant SFH. 

    This constraint refers only to the {\it present} star formation rate. As pointed out by 
    O'Connell (\cite{oconnell}) and Rocha-Pinto \& Maciel (\cite{RPM97}), it is not the same as 
    the star formation {\it history}.

    A typical criticism to a plot like that shown in Figure \ref{sfrh} is that the 
    results still do not rule out a constant SFH, since the oscilations of peaks and 
    lulls around the unity can be understood as fluctuations of a SFH that was `constant' 
    in the mean. This is an usual mistake of those who are accustomed to the 
    strong, short-lived 
    bursts in other galaxies. 

    The ability to find bursts of star 
    formation depends on the resolution. Suppose a galaxy that has experienced 
    only once a real strong 
    star formating burst during its entire lifetime. The burst had an intensity of 
    hundred times the average star formation in this galaxy, and has lasted $10^7$ yr, 
    which are typical parameters of bursts in active galaxies. Figure \ref{probbin} shows 
    how this burst would be noticed, in a plot similar to that we use, as a function 
    of the bin size. In a bin size similar to that used 
    throughout this paper (0.4 Gyr), the strong narrow burst would be seen as a feature 
    with a relative birthrate of 3.5. If we were to convolve it with the age errors, like 
    those we used in Paper I, we could find a broad smeared peak similar to those 
    in Figure \ref{sfrh}. For a biggest bin size (1 Gyr), the relative birthrate 
    of the burst would be lower than 1.5. Hence, a relative 
    birthrate of 1.5 in a SFH binned by 1 Gyr is by no means constant. A great bin size 
    can just hide a real burst that, if occurring presently in other galaxies, would be 
    accepted with no reserves.

      \begin{figure}
      \resizebox{\hsize}{!}{\includegraphics{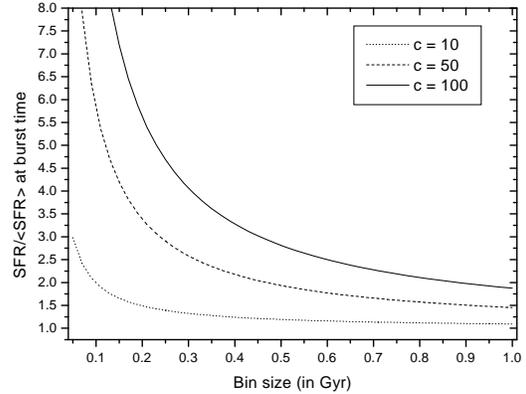}}
      \caption[]{The evanescence of a strong short-lived star formation burst due to the 
        bin size. A star formation burst, lasting 
        $10^7$ yr, and with varying intensity (10, 50 and 100 times more intense than the 
        average star formation rate) was considered. The plot shows the value of the 
        relative birthrate at the time of a burst
        It can be seen that for an age bin similar to that used throughout 
        this paper, namely 0.4 Gyr, even the most strong and narrow burst would be represented 
        by a feature not exceeding 3.5 in units of relative birthrate.}
      \label{probbin}
      \end{figure}

    In the case of our galaxy, the bin size presently cannot be smaller than 0.4 Gyr. 
    This is caused by the magnitude of the age errors. We are then limited to features whose 
    relative birthrate will be barely greater than 3.0, especially taking into consideration 
    that the star formation in a spiral galaxy is more or less well distributed during its 
    lifetime. Therefore, in a plot with bin size of 0.4 Gyr, relative birthrates of 2.0 are 
    in fact big events of star formation. 

    A conclusive way to avoid these mistakes is to calculate the expected fluctuations of a 
    constant SFH in the plots we are using. We have calculated the Poisson deviations 
    for a constant SFH composed by 552 stars. 
    In Figure \ref{sfrerror} we show the 2$\sigma$ 
    lines (dotted lines) limiting the expected statistical fluctuations of a constant SFH. 

    The Milky Way SFH, in this Figure, is presented with two sets of error bars, 
    corresponding to extreme cases. The smallest error bars correspond to Poisson 
    errors ($\pm\sqrt{N}$, where $N$ is the number of stars in each metallicity 
    bin). The thinner longer error bar superposed on the first shows the maximum 
    expected error in the SFH, coming from the combination of counting errors, IMF 
    errors and scale height errors. These last two errors were estimated 
    from Figures \ref{epsilon} and \ref{scaleheight}. The contribution of the scale height  
    errors are greatest at an age of 3.0 Gyr, due to the steep increase of the 
    scale heights around solar-mass stars. The effect of the IMF errors 
    are the smallest, but grows in importance for the older age bins. 

    From the comparison of the maximum expected fluctuations of a constant SFH and the errors in 
    the Milky Way SFH, it is evident that some trends are not consistent with a constant history,  
    particularly bursts A and B, and the AB gap. We can conclude that the irregularities of our 
    SFH cannot be caused by statistical fluctuations.

      \begin{figure}
      \resizebox{\hsize}{!}{\includegraphics{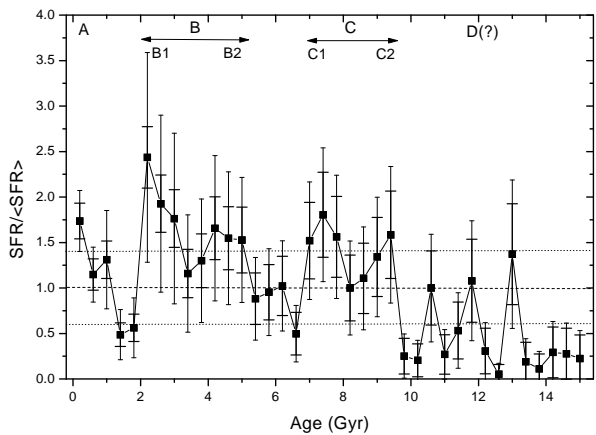}}
      \caption[]{Star formation rate with counting errors. The error bars correspond 
        to an error of $\pm \sqrt N$, where $N$ is the number of stars found in 
        each age bin. The dotted lines indicate the 2$\sigma$ variations around 
        a constant SFR for a sample having 552 stars. The labels over the peaks are 
        the same as in Figure \ref{sfrh}.}
      \label{sfrerror}
      \end{figure}

   \subsection{The uncertainty introduced by the age errors} 

    The age error affects more considerably the duration of the star formation 
    events, since they tend to scatter the stars originally born in a 
    burst. We can expect that this error could smear out peaks and 
    fill in gaps in the age distribution. A detailed and realistic investigation 
    of the statistical meaning of our bursts has to be done in the framework of 
    our method, following the observational data as closely as possible. In the case of 
    the Milky Way, the input data is provided by the age distribution. We have 
    supposed that this age distribution is depopulated from old objects, since some 
    have died or left the galactic plane. Our method to find the SFH makes use of 
    corrections to take into account these effects. However, some features in the 
    age distribution could be caused rather by the incompleteness of the sample. These 
    would propagate to the SFH giving rise to features that could be taken as real, 
    when they are not.

    Thus, if we want to differentiate our SFH from a constant one, we must begin with 
    age distributions, generated by a constant SFH, depopulated in the same way 
    that the Galactic age distribution. With this approach, we can check if 
    the SFH presented in Figure \ref{sfrh} can be produced by errors in the isochrone 
    ages in conjunction with statistical fluctuations of an originally constant SFH.

    We have done a set with 6000 simulations to study this. Each simulation 
    was composed by the following steps:
    \begin{enumerate}
    \item A constant SFH composed by 3000 `stars' was built by randomly 
      distributing the stars from 0 to 16 Gyr with 
      uniform probability.
    \item The stars are binned at 0.2 Gyr intervals. For each bin, we 
      calculate the number of objects expected to have left the main sequence or 
      the galactic plane. This corresponds to the number of objects which we have 
      randomly eliminated from each age bin. The remaining stars (around 
      600-700 stars at each simulation) were put into an `observed catalogue'.
    \item The real age of the stars in the `observed catalogue' is shifted 
      randomly according to the average errors presented in Figure 5 of 
      Paper I. After that, the `observed catalogue' looks more 
      similar to the real data.
    \item The SFH is then calculated just as it was done for the disk. 
      From each SFH the following information is extracted: dispersion around the 
      mean, amplitude and age of occurrence of the most prominent peak, amplitude 
      and age of occurrence of the deepest lull.
    \end{enumerate}

      \begin{figure*}
      \resizebox{\hsize}{!}{\includegraphics{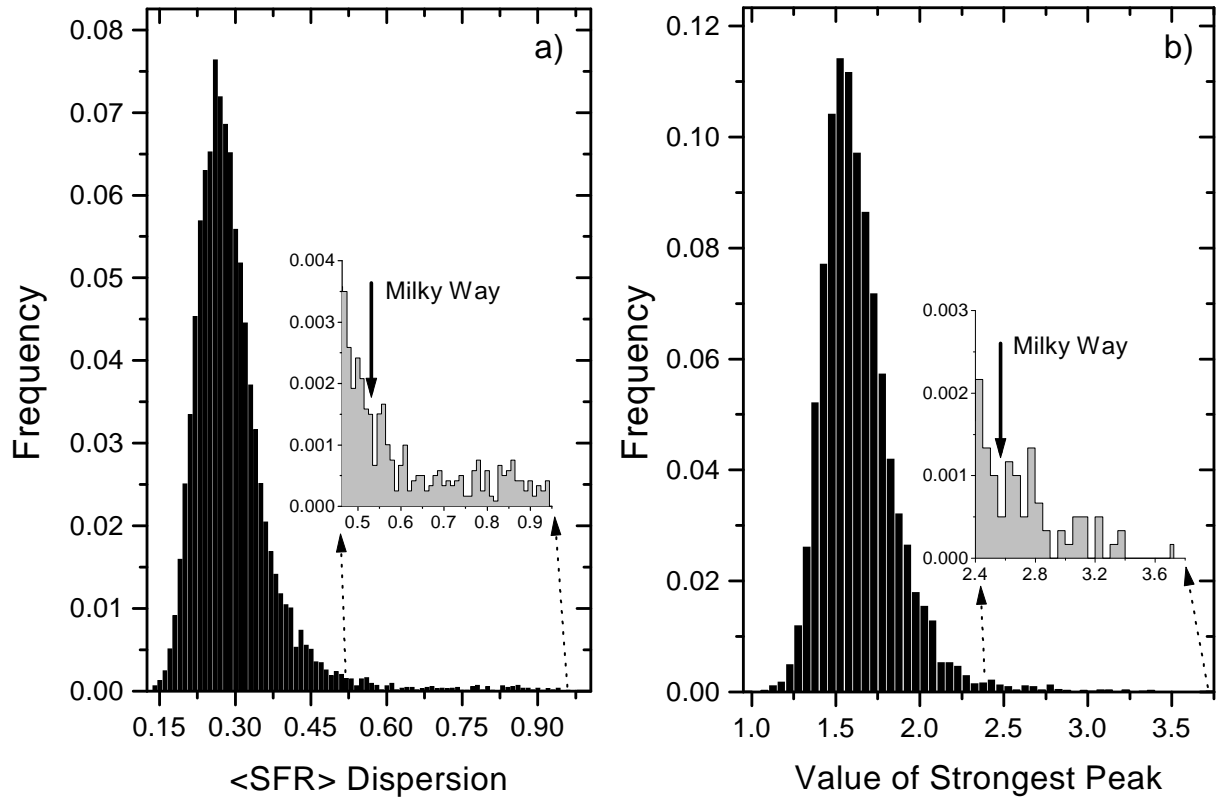}}
      \caption[]{Distribution of parameters from 6000 simulations, 
          using the scale heights from Scalo (\cite{scalo}). The left panel 
          shows the dispersions around the mean SFH, while the right panel gives 
          the value of the most prominent peak. In all the plots, the arrow 
          indicates the corresponding value for the Milky Way SFH.}
      \label{probability}
      \end{figure*}

    One of the problems that we have found is that due to the size of the sample, and 
    the depopulation caused by stellar evolution and scale height effects, the SFH 
    always presents large fluctuations beyond 10 Gyr. These fluctuations are by no means 
    real. They arise from the fact that in the observed sample (for the case of the 
    simulations, in the `observed catalogue'), beyond 10 Gyr, the number of objects 
    in the sample is very small, varying from 0 to 2 stars at most. In the method 
    presented in the subsections above, we multiply the number of stars present in 
    the older age bins by some factors to find the number of stars originally born at 
    that time. This multiplying factor increases with age and could be as high as 
    12 for stars older than 10 Gyr; this way, by a simple statistical effect of 
    small numbers, we can in our sample find age bins where no star was observed 
    neighbouring bins where there are one or more stars. And, in the recovered SFH, this 
    age bin will still present zero stars, but the neighbouring bins would have their 
    original number of stars multiplied by a factor of 12. This introduces large 
    fluctuations at older age bins, so that all statistical parameters 
    of the simulated SFHs were calculated only from ages 0 to 10 Gyr.

    In Figure 
    \ref{probability}, we present two histograms with the statistical parameters 
    extracted from the simulations. The first panel shows the distribution of 
    dispersions around the mean for the 6000 simulations. The arrow indicates the 
    corresponding value for the Milky Way SFH. The dispersion of the SFH of our Galaxy is 
    located in the farthest tail of the dispersion distribution. The probability 
    of finding a dispersion similar to that of the Milky Way is lower than 1.7\%, 
    according to the plot. In other words, we can say, with 
    a significance level of 98.3\%, that the Milky Way SFH is not consistent with a 
    constant SFH.

    In panel b of Figure \ref{probability}, a similar histogram is presented, now for 
    the value of the most prominent peak that was found in each simulation. 
    In the case of the Milky 
    Way, we have B1 peak with $b=2.5$. Just like the previous case, it is also located 
    in the tail of the distribution. From the comparison with the values of the 
    highest peaks that could be caused by errors in the recovering of an originally 
    constant SFH, we can conclude with a significance level of 99.5\% that our 
    Galaxy has not had a constant SFH.

      \begin{figure*}
      \resizebox{\hsize}{!}{\includegraphics{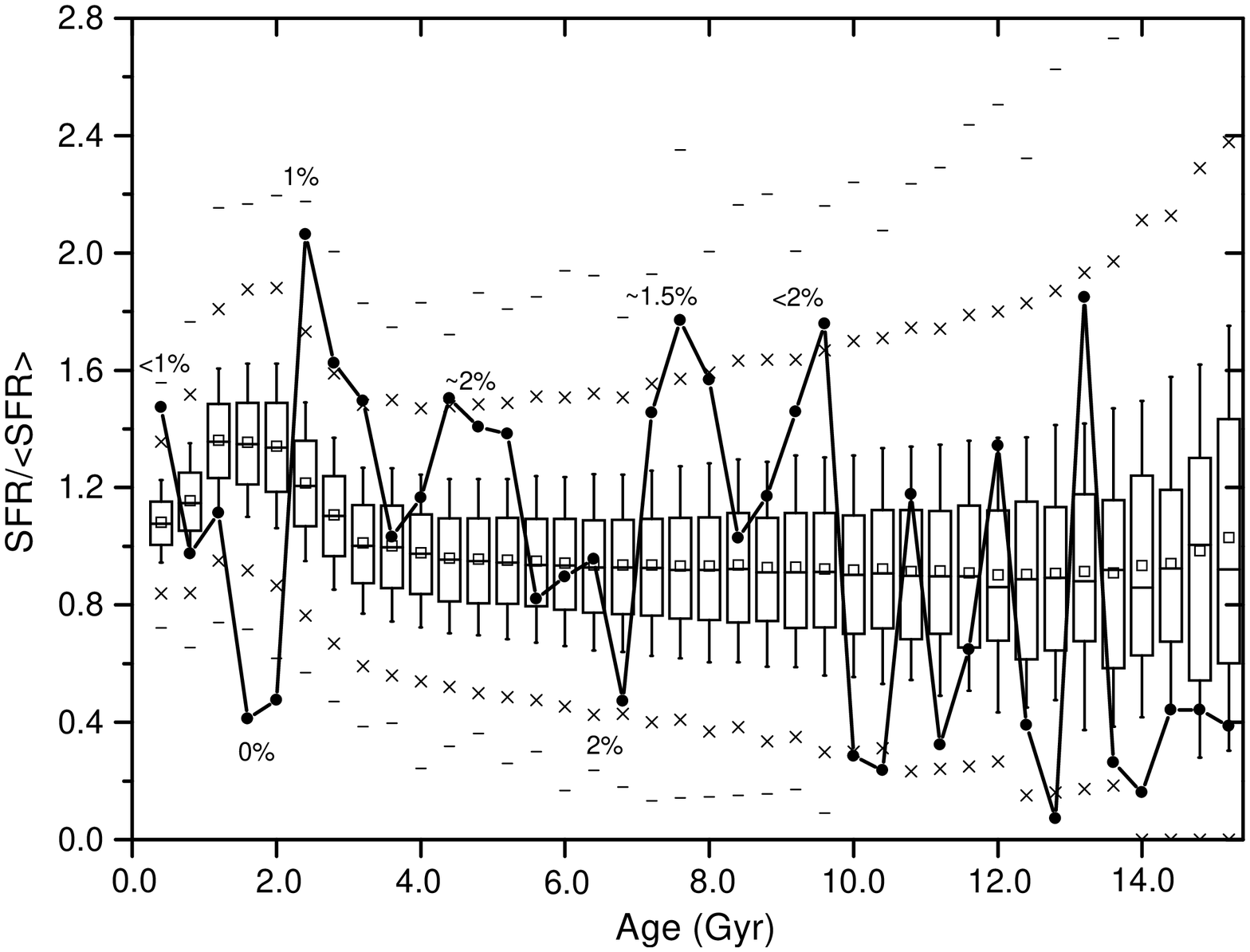}}
      \caption[]{Box charts showing the results of the 6000 simulations, using Holmberg 
          \& Flynn (\cite{holmberg})'s scaleheights. The horizontal lines in the box give   
          the 25th, 50th, and 75th percentile values.  The error bars give the 5th and 95th 
          percentile values.  The two symbols below the 5th percentile error bar give the 
          0th and 1st percentile values.  The two symbols above the 95th percentile error bar 
          give the 99th and 100th percentiles.  The square symbol in the box shows the mean 
          of the data. Superimposed, the Milky Way SFH is shown. From the comparison with 
          the distribution of results at each age bin, the probability to find each particular 
          event in a constant SFH can be calculated. The numbers besides the major events 
          give the probabilities for their being fluctuations of a constant SFH.}
      \label{allchance}
      \end{figure*}

    The use of Holmberg \& Flynn (\cite{holmberg}) scale heights in the simulations 
    increases these  
    significance levels to 100\% and 99.9\%, respectively.

    These significance levels refer to only one parameter of the SFH, namely 
    the dispersion or the highest peak. For a rigorous estimate of the probability 
    of finding a SFH like that presented in Figure \ref{sfrerror}, from an originally 
    constant SFH, one has to calculate the probability to have neighbouring bins with 
    high star formation, followed by bins with low star formation, as a function of age. 
    This can be calculated approximately from Figure \ref{allchance}, where we show 
    box charts with the results of the 6000 simulations. Superimposed on these box charts, 
    we show the SFH, now calculated with Holmberg \& Flynn (\cite{holmberg})'s 
    scale heights. For the sake of consistency, the simulations shown in the figure also 
    use these scale heights, but we stress that the same quantitative result is found 
    using Scalo's scale heights.

    A lot of information can be drawn from this figure. First, it can be seen that a 
    typical constant SFH would not be recovered as an exactly `constant' function in 
    this method. This is shown by the boxes with the error bars which delineate 
    2$\sigma$-analogous to those lines shown in Figure \ref{sfrerror}. The boxes 
    distribute around unity, but shows a bump between 1 to 2 Gyr, where the average 
    relative birthrate increases to 1.4. This is an artifact introduced by the age 
    errors. In each individual simulation, the number of stars scattered off their 
    real ages increases as a function of the age. In the recovered SFH there will be 
    a substantial loss of stars with ages greater than 15 Gyr, since they are eliminated 
    from the sample (note that originally, these stars would present ages lower than 
    15 Gyr, and just after the incorporation of the age errors they resemble stars 
    older than 
    it). This decreases the average star formation rate with respect to the original SFH, 
    and the proportional number of young stars increases, because they are less 
    scattered in age due to errors. This gives rise to a distortion in the expected 
    loci of constant SFHs. Note also the increase in the $2\sigma$-region as we 
    go towards older ages, reflecting the growing uncertainty of the chromospheric ages. 

    The diagram allows a direct estimate of the probability for each feature found in the 
    Milky Way SFH be produced by fluctuations of a constant SFH. The box charts gives the 
    distribution of relative birthrates in each age bin. An average probability for 
    the major events of our SFH are shown in Figure \ref{allchance}, besides the 
    features under interest. Rigorously speaking, 
    the probability for the whole Milky Way SFH be constant, not bursty, can be estimated by 
    the multiplication of the probability of the individual events in this Figure. It 
    can be clearly seen that it is much less than the 2\% level we have calculated from only 
    one parameter of the SFH. Particularly, note that the AB gap has zero probability to be 
    caused by a statistical fluctuation. 
    All of theses results show that the Milky Way SFH was by no means constant.

  \subsection{Flattening and Broadening of the Bursts}

    Since the errors in the chromospheric ages are not negligible, a sort of smearing 
    out must be present in the data. Due to this, a star formation burst found in the 
    recovered SFH must have been originally much more pronounced. This mechanism probably 
    affects much more older bursts, since the age errors are greater at older ages 
    and the depopulation by evolutionary and scaleheight effects 
    is more dramatic. 
    We can assume that if we found a feature like a burst at say 8 Gyr ago, this 
    probably was much stronger in order to be preserved in the recovered SFH.

    The first aspect we want to show is that the errors produce a significant 
    flattening of the original peaks. To do so, we use simulations of a SFH composed 
    by a single burst over a constant star formation rate. The `burst' is characterized 
    by occurring at age $\tau$, having intensity $c$ times the value of the constant 
    star formation rate, and lasting 1 Gyr. We want to know the fraction of the burst 
    that is recovered, as a function of age and of the burst intensity.

    We have performed 50 simulations for each pair $(\tau,c)$, with around 3000 stars in each 
    simulation. A summary of these simulations is shown in Figure \ref{smearsummary}. 
    In all the 
    panels (for varying $c$), the fraction of the recovered burst is high for recent 
    bursts and falls off smoothly until 8-9 Gyr, when it begins to become constant. 
    This stabilization reflects the predominance of the statistical fluctuations, 
    since the recovered fraction is the same, regardless of the age of occurrence. What 
    happens is that the burst becomes more or less undistinguished from the fluctuations. 
    From this we can conclude that it is more difficult to find bursts older than 8-9 
    Gyr, irrespective of its original amplitude.

      \begin{figure*}
      \resizebox{\hsize}{!}{\includegraphics{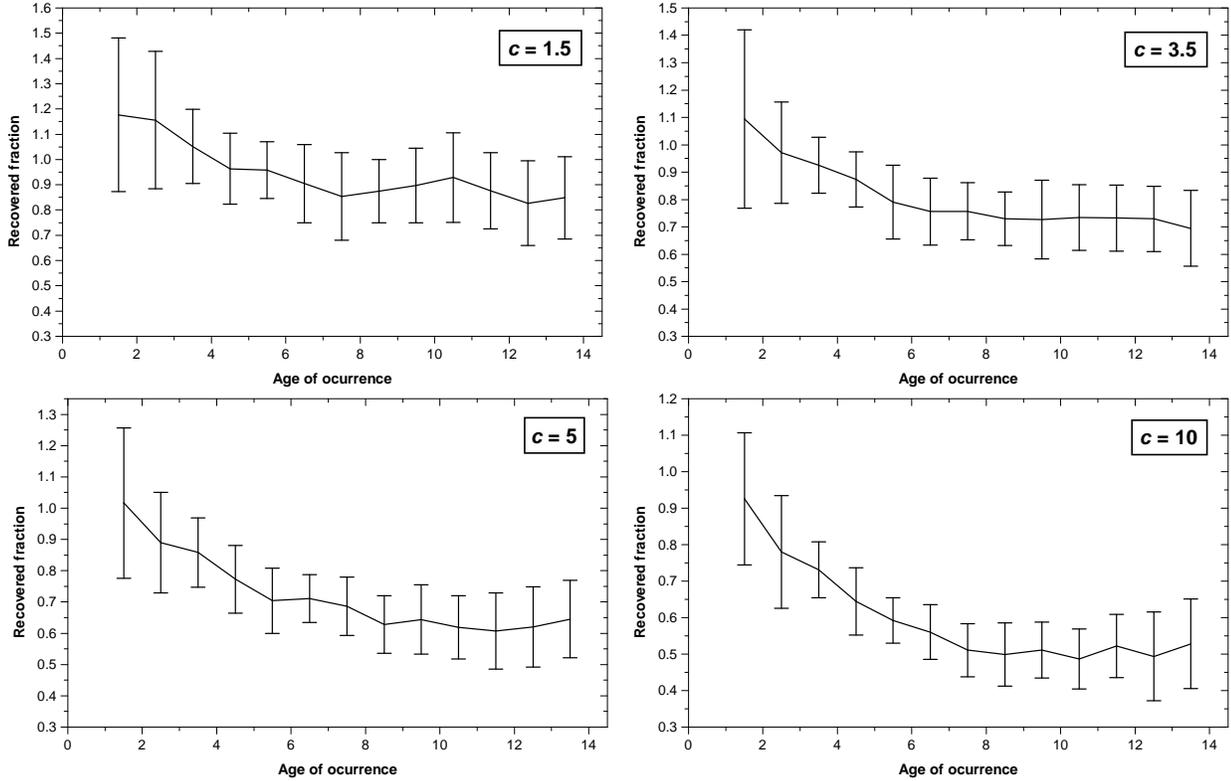}}
         \caption{Recovered fractions for a SFH composed of a 
             single burst superimposed on a constant rate. The `burst' is characterized 
             by occuring at age $\tau$, having intensity $c$ times the value of the 
             constant star formation rate, and lasting 1 Gyr. We show the cases for 
             $c = $ 1.5, 3.5, 5 and 10. In all the plots, the abscissa indicates the 
             age $\tau$ where the burst happened. The fraction recovered in the first 2 
             Gyr of age is greater than unity, due to the same problem that distorted 
             the $2\sigma$ loci of the constant SFHs in Figure \ref{allchance} (see text).}
      \label{smearsummary}
      \end{figure*}

    A second problem in the method is the broadening of the bursts. This depends sensitively 
    on the age at which the burst occurs, and the results are even more dramatic. To 
    illustrate this, another set of simulations was done. We consider now a SFH composed 
    of a single burst, of 1000 stars, lasting 0.4 Gyr. No star formation occurs except 
    during the burst. We vary the age of occurrence from 0.3 Gyr to 6 Gyr ago. Just one 
    simulation was done for each age of occurrence, since we are only looking for the magnitude 
    of the broadening introduced by the errors, so the exact shape of the recovered 
    SFH does not matter. The recovered 
    SFHs are shown in Figure \ref{broad}. Only the younger bursts 
    are reasonably recovered. The burst at 6 Gyr can still be seen, although many of its 
    stars has been scattered over a large range of ages.

      \begin{figure}
      \resizebox{\hsize}{!}{\includegraphics{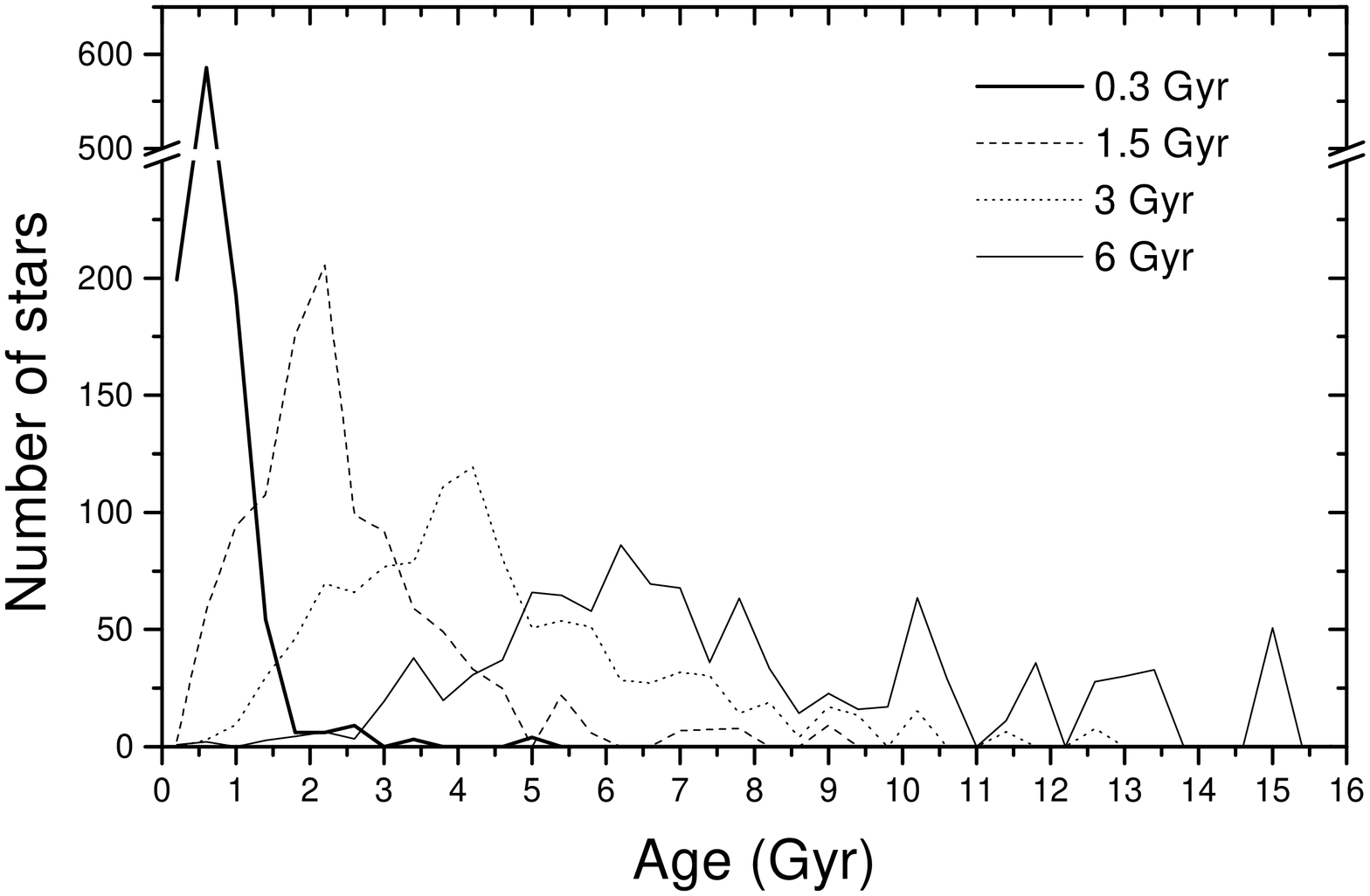}}
         \caption{Recovered SFHs for an original SFH composed of a single burst of 1000 
           stars. The curves show examples of how these bursts are broadened, depending on 
           their age of occurrence, due to age errors.}
      \label{broad}
      \end{figure}

    \section{Comparison with other constraints}

    \subsection{The SFH driving the chemical enrichment of the disk}

    On theoretical grounds, there should be a correlation between the SFH and the 
    the age--metallicity relation (hereafter AMR). The increase in the star formation 
    leads to an 
    increase in the rate at which new metals are produced and ejected into the interstellar 
    medium. The correlation is not a one-to-one, since the presence of infall and radial 
    flows can also affect the enrichment rate of the system. Moreover, the enrichment 
    rate is constrained by the amount of gas into which the new metals will be diluted. 
    Nevertheless, it is interesting to see whether the AMR we have found in Paper I 
    is consistent with the SFH derived from the same sample, especially because, to our 
    knowledge, this was never tried before.

    From the basic chemical evolution equations (Tinsley \cite{tinsley}), for a closed 
    box model (i.e., no infall), the link between the AMR and the SFH can be written as 
    \begin{equation}
      {{\rm d}Z\over{\rm d}t}(t)\propto{\psi(t)\over m_g(t)},
      \label{chemo}
    \end{equation}
    where $Z(t)$ gives the AMR, expressed by absolute metallicity, $\psi(t)$ is the 
    SFH as in equation (\ref{Nast}), and $m_g(t)$ is the total gas mass of the system, 
    in units of $M_\odot$~pc$^{-2}$. 

      \begin{figure}
      \resizebox{\hsize}{!}{\includegraphics{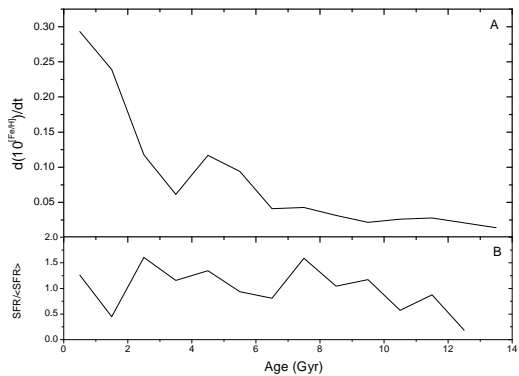}}
      \caption[]{Comparison between the metal-enrichment rate (defined as the derivate in time 
          of the AMR, expressed by absolute metallicity $Z$), and the SFH. On theoretical 
          grounds both quantities should be correlated, and it should give a test to the 
          reliability of the SFH. In practice, the magnitude of the errors in both 
          functions still hinders the application of this test.}
      \label{chemoburst}
      \end{figure}

    According to this equation, bursts in the SFH are echoed through an increase of the 
    metal-enrichment rate. Certainly, this is particularly true when the metallicity is 
    measured by an element produced mostly in type II supernovae, like O. 
    The gas mass can dilute more or less the enrichment, changing the 
    proportionality between it and the SFH, at each age, but will not destroy the relationship. 
    On the other hand, the intrinsic metallicity dispersion of the interstellar medium can 
    certainly somewhat obscure this proportionality, especially if it were as big as the 
    AMR by Edvardsson et al. (\cite{Edv}, 
    hereafter Edv93) suggests.

    In Figure \ref{chemoburst}, we show a comparison between the 
    metal-enrichment rate (top panel) with the SFH (bottom panel). The enrichment rate 
    increases substantially in the last 2 Gyr, which could be a suggestion for a recent 
    burst of SFH. However, the agreement between both functions seems very poor. There 
    is a peculiar bump in the enrichment rate between 4 and 6 Gyr, which is coeval to 
    a feature in the SFH, but most probably this is mere coincidence. 

    Although we have used iron as a metallicity indicator, which invalidates  
    Equation (\ref{chemo}), due to non recycling effects, we are not sure whether the 
    situation would be improved by using O. The errors in both the AMR and SFH are still 
    big enough to render such a comparison extremely uncertain. 
    However, it can be a test to be done with improved data. 
    The more important result for 
    chemical evolution studies is that, provided that we know accurately both functions, 
    the empirical AMR and SFH will allow an estimate of the variation of the gas mass with 
    time, which could lead to an estimate of the evolution of the infall rate. Future studies 
    should attempt to explore this tool.

  \subsection{Scale length of the SFH}

    The stars in our sample are all presently situated within a small volume of 
    about 100 pc radius around the Sun. The star formation history derived from 
    these stars is nevertheless applicable to a quite wide section of the 
    Galactic disk, since the stars which are presently in the Solar 
    neighbourhood have mostly arrived at their present positions from a torus in 
    the disk concentric with the Solar circle.

    We have investigated how wide this section of disk is by integrating the 
    equations of motion for 361 stars of the `kinematic sample' (see Paper I) 
    within a model of the Galactic potential. 
    The potential consists of a thin exponential disk, a spherical bulge and a 
    dark halo, and is described in detail in Flynn et al. (\cite{flynn1996}). 
    For each star we 
    determine the orbit by numerical integration, and measure the peri- and 
    apogalactic distances, $R_p$ and $R_a$ and the mean Galactocentric radius, 
    $R_m = (R_p + R_a)/2$ for the orbit (cf. Edvardsson et al. \cite{Edv}). 

    The distribution of $R_m$ is shown in Figure \ref{orbits}. Most of the stellar orbits 
    have mean Galactocentric radii within 2 kpc of the Sun (here taken to be at 
    $R_\odot = 8$ kpc), i.e. $6 < R_m < 10$ kpc. Very few stars in the sample 
    are presently moving along orbits with mean radii beyond these limits.

    As discussed by Wielen, Fuchs and Dettbarn (\cite{WFD}), due to irregularities in 
    the Galactic potential caused by (for example) giant molecular clouds and 
    spiral arms, the present mean Galactocentric radius of a stellar orbit 
    $R_m(t)$ at time $t$ does not bear a simple relationship to the mean 
    Galactocentric radius of the orbit on which the star was born $R_m(0)$. 
    Wielen, Fuchs and Dettbarn describe the process by which stars are scattered 
    by these irregularities as orbital diffusion, and show that over time scales 
    of several Gyr, that one cannot reconstruct from $R_m$ the radius at which 
    any particular star was born to better than a few kpc. This is of the same 
    order as the width of the distribution of $R_m$ seen in Figure \ref{orbits}. We 
    therefore conclude that our stars fairly represent the star formation 
    history within a few kpc of the present Solar radius, $6 < R_m < 10$, or the 
    ``middle distance'' regions of the Galactic disc. The SFH of the 
    inner-disk/bulge, and the outer disk are not sampled. 

      \begin{figure}
      \resizebox{\hsize}{!}{\includegraphics{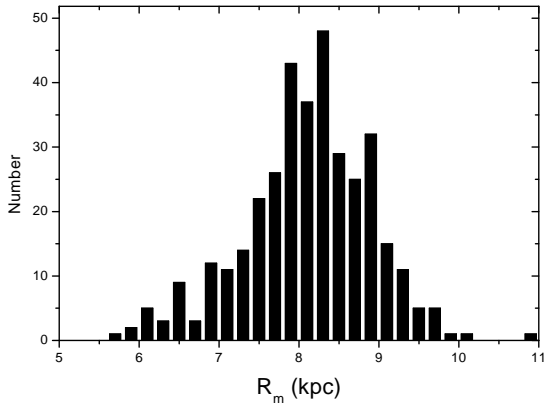}}
      \caption[]{The histogram presents the mean galactocentric radius for the orbits of 
         361 stars presently located near the Sun. According to Binney \& Sellwood (2000), 
         these mean galactocentric radii indicate with good confidence the stellar birthplace 
         radii. This shows that the SFH here derived is not a local history, but the average 
         history over a significant part of the Milky Way disk.}
      \label{orbits}
      \end{figure}

    However, Binney \& Sellwood (\cite{binney}) have criticized this conclusion. They show that during 
    the lifetime of a star, the guiding-center of its orbit can change generally by no more than 5\%. 
    In this scenario, the value of $R_m$ that we have calculated is close to the galactocentric 
    radius of the star birthplace, and our star formation history would still be representative of a 
    considerable fraction of the galactic disk, $7 < R_m < 9$. 

    Another important conclusion of kinematic studies it that the older is a feature 
    in the SFH, the more damped it is  
    recovered from the data, related to its original amplitude (see, for example, 
    Meusinger \cite{meus}), since the stars formed by the burst will be scattered through 
    a larger region. Hence, the younger bursts in our SFH are the most local features. This does 
    not mean that they are most probably `local irregularities'. In time scales of 1-2 Gyr, the 
    diffusion of stellar orbits homogenize any irregularities in the azimutal direction, so 
    that the bursts would apply to the whole solar galactocentric annulus.

  \subsection{The Galaxy and the Magellanic Clouds}

    When evidences for an intermittent SFH in the Galaxy were first discovered, Scalo 
    (\cite{scalo87}) proposed that they could have originated from interactions between 
    the Galaxy and the Magellanic Clouds. 
    Indeed, the Magellanic Clouds are known to have probably experienced some episodes 
    of strong star formation for a long time. Butcher (\cite{acougueiro}) first proposed 
    that the bulk of star formation in the 
    Large Magellanic Cloud (LMC) has occurred from 3-5 Gyr ago, by the 
    analysis of the luminosity function of field stars. Stryker et 
    al. (\cite{stryker1}) and Stryker (\cite{stryker2}) subsequently confirmed 
    this result. In the last few years, additional studies have arrived almost at the 
    same conclusions (Bertelli et al. \cite{bertelli}; Vallenari et al. 
    \cite{vallenari1},b}). 
    Westerlund (\cite{wester}) also remarked that the star formation in the LMC 
    seems to have been very small from 0.7 to 2 Gyr ago. A very recent burst of 
    star formation (around 150 Myr ago) was also found by the MACHO team (Alcock 
    et al. \cite{todopau}) from the study of the period distribution of 1800 
    LMC cepheids. Their analysis present compeling arguments favouring this 
    hypothesis, as well as for the propagation of the star formation to neighbour 
    regions.

    However, these results have more recently been questioned, on the basis of 
    colour-magnitude diagram synthesis. Some authors 
    claim that important information on the SFH are provided by 
    the part of the colour--magnitude diagram below the turnoff-mass, which 
    could only be resolved with the most recent observations (Holtzman 
    et al. \cite{vera_holtz}, \cite{holtz99}, and references therein; 
    Olsen \cite{xxx}). These papers conclude that star 
    formation in the LMC has been a continuous process over much of its 
    lifetime. 

    Note that {\it continuity} in the SFH does not means {\it constancy}. Holtzman 
    et al. (\cite{holtz99}) points that their method cannot constrain accurately the 
    burstiness of the SFH in the LMC on small time scales, particularly for ages 
    greater than 4 Gyr. Nevertheless, they show evidence for an increase in the 
    star formation rate in the last 2.5 Gyr. Dolphin (\cite{dolphin}) arrives to 
    the same conclusion studying two different fields of the LMC, separated by around 
    2 kpc one from the other. The author recognizes that some large environment 
    alteration must have triggered an era of star formation in our neighbour galaxy. 

    In spite of the controversy, it is impossible not to verify that some 
    results on the SFH of the LMC are in 
    apparently synchronism with some SFH events in the Milky Way disk. But this 
    should be not really surprising. The Magellanic Clouds are satellites of our Galaxy, and  
    past interactions between them were a rule, not an exception. Byrd \& 
    Howard (\cite{byrd}) showed that a companion satellite, whose mass is larger than 
    1\% of the primary galaxy, could excite large-scale tidal arms in the disk of 
    the primary, and we know that spiral arms do induce, or at least organize, 
    star formation. This number is 
    to be compared with the mass ratio between our Galaxy and the 
    Clouds which is 0.20 (Byrd et al. \cite{byrdetal}). Besides direct tidal effects, the 
    Clouds can produce a dynamical wake in the halo that distorts the disk (Weinberg \cite{weinberg}). 
    It is quite possible that such an effect could also enhance the star formation in the 
    disk (M. Weinberg, private communication).
    
    Additional evidence comes from dynamical studies of the Magellanic Clouds. Several 
    groups have worked on the derivation of their orbits around the Galaxy. The full 
    orbit of the Magellanic Clouds are still unknown, but there is some agreement 
    in the published works. The most 
    important is that all of these works conclude that the most recent close encounter 
    between the Clouds and the Milky Way has occurred 0.2-0.5 Gyr ago, which was the 
    closest encounter through the entire history of the system (however, Holtzman et 
    al. \cite{holtz99} mention an unpublished work by Zhao in which the last 
    perigalacticon occurred 2.5 Gyr ago). 
    Murai \& Fujimoto 
    (\cite{murai}) calculated that other close encounters have occurred 1.5, 2.6 and 
    7.5 Gyr ago. Gardiner et al. (\cite{gardiner}) revisited Murai \& Fujimoto 
    (\cite{murai})'s model and recalculated the epochs of the close encounters as 
    around 1.6, 3.4, 5.5, 7.6 and 10 Gyr ago. However, Lin et al. (\cite{lin}) have 
    found different values: 2.6, 5.3, 8.4 and 11.8 Gyr ago.

    From these results we can tentatively assume that, in the last 12 Gyr, the Clouds 
    have had at most six close encounters with the Milky Way occurring more or less at 
    0.2-0.5, 1.4-1.5, 2.6-3.4, 5.3-5.5, 7.5-8.4 and 10-11.8 Gyr ago. Some of these 
    encounters are 
    not predicted by all the authors, while some are in good agreement. For the 
    sake of simplicity, we will refer to these encounters as I, II, III, IV, V and VI, 
    respectively. 

\begin{figure}
      \resizebox{\hsize}{!}{\includegraphics{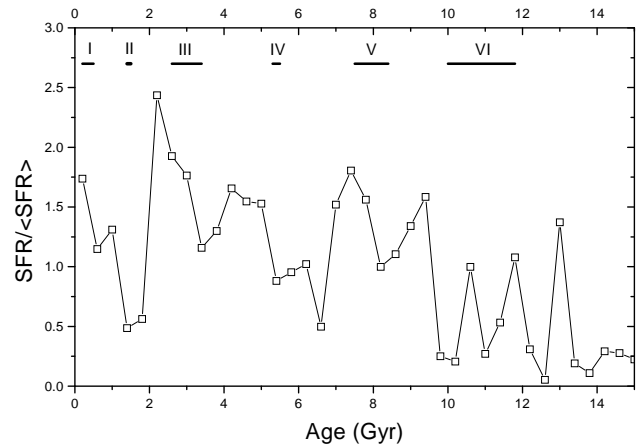}}
      \caption[]{Star formation history compared with the times of close encounters 
           between the Galaxy and the Magellanic Clouds.}
      \label{magellan}
      \end{figure}

    There are similarities 
    between the time of close encounters and 
    the events of our derived SFH. In Figure \ref{magellan} we show the epoch of 
    these encounters superimposed over our SFH. We can associate burst A with 
    encounter I, peak B1 with encounter III, and peak C1 with encounter V. It is not 
    unlikely that peak B2 could also be associated with encounter IV. On the other 
    hand, encounter VI probably cannot be responsible for any of the features 
    found beyond 9 Gyr, since it occurs in an age range where the SFH is highly uncertain 
    and subject to random fluctuations. 

    A significant exception to the rule is encounter II. It is thought to have happened 
    in the middle 
    of the AB gap. It seems strange to think that a close encounter between interacting 
    galaxies could suppress the star formation. Other mechanism should be 
    responsible for the gap. On the other hand, Lin et al. (\cite{lin}) have not 
    found such an encounter. In fact, these authors predict that by this time, the 
    Clouds would be located in their apogalacticon, more than 100 kpc away.  

    Although the comparison is very premature, we conclude that 
    the data on the age distribution and 
    orbits of the Magellanic Clouds present some agreement with the 
    Miky Way SFH. Have the bursts of star formation in the Milky Way 
    been produced by interaction with its satellite galaxies? The comparison 
    above certainly points to this possibility, that deserves more 
    investigations to be properly answered, since there is still much 
    uncertainty in the Magellanic Clouds close encounters, as well as on the 
    chronologic scale of the chromospheric ages.

    \section{The features of the Milky Way SFH}

    We now can return to the discussion of the meaning of each feature found in the SFH derived 
    in section 3.

    \subsection{Burst A}
    
    The most recent star formation burst is also the most likely burst to have occurred,  
    since it has 
    occurred in the very recent past, and so is less affected by the age errors. A 
    recent enhancement in the SFH is also present in nearly all previous investigations of 
    the SFH (Scalo \cite{scalo87}; Barry \cite{barry}; G\'omez et al. \cite{gomez}; 
    Noh \& Scalo \cite{noh}; Soderblom et al. \cite{soder91}; Micela et al. \cite{micela}; 
    Rocha-Pinto \& Maciel \cite{RPM97}; Chereul et al. \cite{chereul}), and is consistent 
    with the distribution of spectral types in class V stars (Vereshchagin \& Chupina 
    \cite{veresh}). It is not present in the 
    isochrone age distributions (Twarog \cite{twar}; Meusinger \cite{meusinger}) most 
    probably due to the difficulty to measure ages for stars near the zero-age main sequence, 
    where we expect to find the components of this burst in a HR diagram. 

    We can conclude with confidence 
    that it is a real feature of the SFH. However, 
    being the youngest, it is also the most {\it local} feature, because the younger stars have 
    had no time to diffuse to larger distances from their birthsites. Thus, we cannot be sure  
    (from out data only) whether this feature applies to the Milky Way as a whole. 
  
    On the other hand, it is known that the Large Magellanic 
    Cloud appears to have experienced also a recent burst of star formation (Westerlund 
    \cite{wester}; Alcock et al. \cite{todopau}) which is very well represented by its 
    young population of open 
    clusters, cepheids, OB associations and red supergiants. At the time of this burst, both 
    galaxies have been closer than ever in their history (Lin et al. \cite{lin}). 
    This suggests that burst A could be 
    caused by tidal interactions between our Galaxy and the LMC.

    \subsection{AB gap}

    A substantial depression in the star formation rate 1-2 Gyr ago was found by 
    many studies, beginning with Barry (\cite{barry}; see also the SFH derived from the 
    massive white dwarf luminosity function derived by Isern et al. \cite{isern}). This gap 
    appears, although not directly, in the chromospheric age distribution 
    (the so-called Vaughan-Preston gap) and 
    in the spectral type distribution, between A and F dwarfs (Vereshchagin 
    \& Chupina \cite{veresh}). A quiescence 
    between 1 and 2 Gyr is also visible in Chereul et al. (\cite{chereul}), in their study of the 
    kinematical properties of A and F stars in the solar neighbourhood.

    This feature has been present in 
    all steps of our work, from the initial age distribution in Figure \ref{volume} to 
    the SFH. Note that the volume corrections have deepened this lull, but it has not 
    changed its duration.

    The AB gap is likely to have lasted for a billion years. Previous studies have given 
    a more extended duration for it, but we believe that it was caused by the use of a 
    highly incomplete sample, together with a chromospheric age calibration that 
    does not account for the different chemical composition of the stars.
    Since it is a relatively recent feature, it only samples birthsites over a radial  
    length 
    scale of 1-2 kpc. 

    \subsection{Burst B}

    The small lull between the peaks B1 and B2 is not present in the initial age 
    distribution (Figure \ref{volume}), appearing only after the volume corrections. 
    It is very narrow, which could be most 
    probably caused by hazardous small weights of the stars in these age bins, during 
    the volume correction. This is why we have presently no means to distinguish burst 
    B from a single burst or an unresolved double burst. At its age of occurrence, 
    considerable broadening of the original features is expected.  Either 
    way, our simulations 
    give strong support to this feature. 

\begin{figure}
     \resizebox{\hsize}{!}{\includegraphics{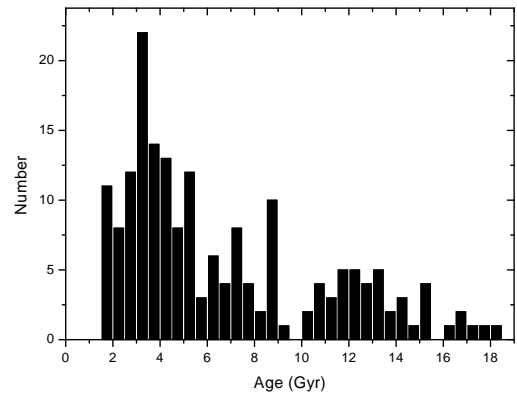}}
    \caption{Age distribution of Edv93's sample. The concentration of stars around 4 Gyr shows evidence 
         of burst B. The lack of objects with ages lower than 2 Gyr is not an evidence of the AB gap, 
         since Edv93 avoided the inclusion of very young stars due to the difficulty to measure an 
         isochrone age for them.}
      \label{edvage}
      \end{figure}

    Previous studies have found star formation 
    enhancements around 4 Gyr ago (Scalo \cite{scalo87}; Barry \cite{barry}; Marsakov et al. 
    \cite{marsakov}; Noh \& Scalo \cite{noh}; Soderblom et al. \cite{soder91}; Twarog \cite{twar}; 
    Meusinger \cite{meusinger}). Note that a strong concentration of stars around this age 
    can also be found in the age distribution of Edv93's stars, that we show in Figure \ref{edvage}. 

\begin{figure}
     \resizebox{\hsize}{!}{\includegraphics{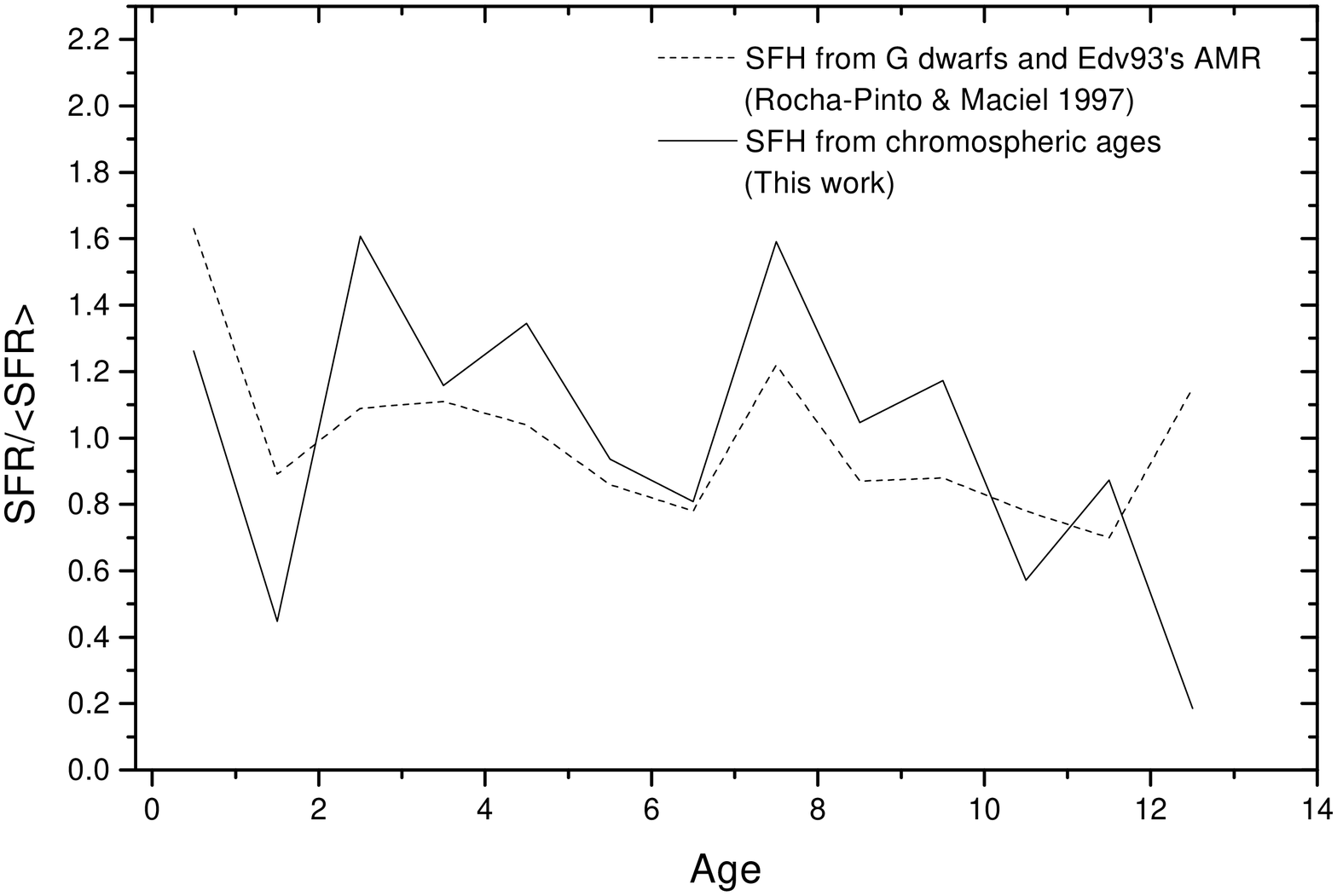}}
    \caption{Comparison between Rocha-Pinto \& Maciel (\cite{RPM97})'s SFH 
           (for Edv93's AMR) and the present chromospherically-based 
           SFR binned by 1 Gyr intervals.}
      \label{we}
      \end{figure}

    A significant exception is the SFH found 
    by some of us (Rocha-Pinto \& Maciel \cite{RPM97}). This paper suggests that burst B 
    would be much smaller than the preceding burst C. To find the SFH, Rocha-Pinto \& 
    Maciel used a method to extract information from the G dwarf metallicity distribution 
    (Rocha-Pinto \& Maciel \cite{RPM96})
    aided by the AMR (see also Prantzos \& Silk \cite{prantzos}). 
    The authors have used several AMRs from the literature and 
    different SFHs were found for each AMR. The SFHs recovered with the AMR from Twarog 
    (\cite{twar}) and Meusinger et al. (\cite{meu}) were preferred compared to that 
    found with Edvardsson et al. \cite{Edv} (hereafter Edv93) AMR. 
    To be consistent with our present result, we need to 
    compare the present SFH with that coming from Rocha-Pinto \& Maciel's method for an AMR 
    similar to that found from our sample (paper I). Our AMR now looks very similar 
    to the mean points 
    of Edv93's AMR. Rocha-Pinto \& Maciel (\cite{RPM97}) have found, using Edv93's AMR, 
    that Burst B could have around the same intensity as burst C, and also a 
    narrow AB gap lasting 1 Gyr at most. Figure \ref{we} shows a comparison between 
    their SFH (for Edv93's AMR) and the present history binned by 1 Gyr intervals. 

      \begin{figure}
     \resizebox{\hsize}{!}{\includegraphics{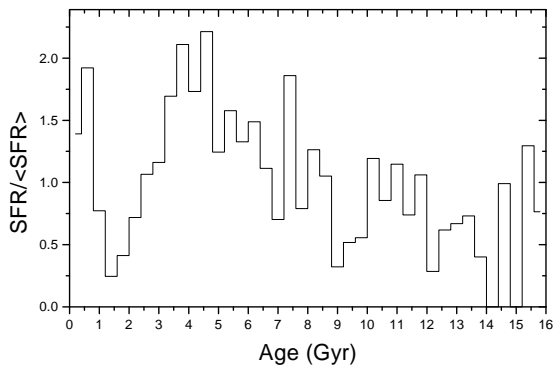}}
         \caption{Results from an example simulation to show how the observed features: 
          Burst A and B and AB gap could be caused by two bursts separated by a lull 
          of star formation. See text for description of the original SFH used in this 
          simulation.}
      \label{tentativa}
      \end{figure}

   \subsection{BC gap and Burst C}

    The existence of the BC gap is directly linked with how much credit we are going to 
    give to Burst C. From Figure \ref{broad}, one could say that no burst could be 
    found around 8-9 Gyr, and all supposed features are artificial patterns created by 
    statistical fluctuations. To reinforce this theoretical expectation, we have done 
    a simulation to show how the features above could be formed 
    by a bursty SFH. We have considered a SFH composed by three bursts, one occurring 
    at 0.3 Gyr, lasting 0.2 Gyr, and the other at 4 Gyr, also lasting 0.2 Gyr, and the last 
    ocurring at 9 Gyr, lasting 0.5 Gyr. The 
    first burst and the last burst are composed by 300 stars, while the second burst 
    is three times more intense. The star formation at other times is assumed to be 
    highly inefficient, forming only 60 more stars at the whole lifetime of the galaxy. 
    The recovered SFR is shown in Figure \ref{tentativa}. Although the two more recent 
    bursts can be well recovered, there is no sign of burst C at 9 Gyr. We have tried other 
    combinations between the amplitude and time of occurrence of them, but in all cases 
    the stars of burst C were much scattered from its original age. 
    
    If on theoretical grounds there is no convincing arguments to accept the existence 
    of burst C, the same does not occur on observational grounds. This puzzling situation 
    comes from the fact that burst C has appeared in a number of studies that have 
    used not only different samples, but also different methods (Barry \cite{barry}; 
    Noh \& Scalo \cite{noh}; Soderblom et al. \cite{soder91}; Twarog \cite{twar}; 
    Meusinger \cite{meusinger}; Rocha-Pinto \& Maciel \cite{RPM97}). And it appears 
    double-peaked in some of them, as we saw in section 3.

    The magnitude of the age errors prevents us from assigning a 
    good statistical confidence 
    to this particular feature. 

    However, it is not implausible that we have overestimated the age errors. A 
    decrease of 0.05 dex in the age errors could alleviate the situation and 
    allow the identification of peaks (although highly broadened) younger than 10 Gyr, which 
    would suggest that burst C is a real feature. 
    A better estimate of the age errors would not create new bursts, or flatten 
    the recovered SFH in these age bins, but would give confidence limits for the ages 
    where the features found are likely to be real and not just artifacts.

  \subsection{Burst D}

    The so-called burst D was proposed by Majewski (\cite{majewski}), as a star 
    formation event that would be responsible for the first stars 
    of the disk, before the formation of the thin disk. 

    A superficial look at Figure \ref{sfrh} could give us the impression that the 
    peaks beyond 11 Gyr were remnants of this predicted burst. However, as we have 
    shown above, it is presently impossible to recover the SFH correctly at this 
    age range, even if our age errors are overestimated by as much as 0.05 dex. 
    The SFH at older ages are dominated by fluctuations, superimposed on 
    the original strongly broadened structures, in such a way that it is imposible 
    to disentangle statistical fluctuations from real star formation events. 
 
    Theoretically, 
    patterns as old as 13 Gyr could be found in the SFH, provided that they occurred not very 
    close to younger ones,  
    if the age errors were decreased by 0.10 dex, but that is hardly possible to 
    be attained at the present moment since it would need to be of the order of 
    magnitude of the error in the $\log R'_{\rm HK}$ index.  

    For these reasons, we give no credit to the peaks beyond 11 Gyr in Figure \ref{sfrh}. 
    If burst D has ever occurred, probably the present chromospheric age distribution 
    is not an efficient tool to find its traces.

\section{The shape of the chromospheric activity--age relation}
    
Soderblom et al. (\cite{soder91}) argued that the interpretation of the chromospheric 
activity distribution as evidence for a non-constant SFH is premature. Particularly, the authors have shown that 
the observations do not rule out a non-monotonic chromospheric activity--age relation, even considering 
that the simplest 
fit to the data is a power-law, like the one we used.

      \begin{figure}
     \resizebox{\hsize}{!}{\includegraphics{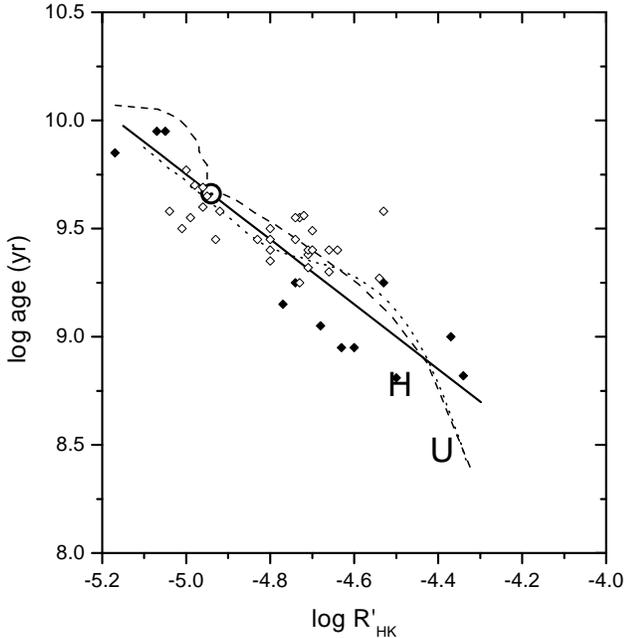}}
         \caption{Comparison between the chromospheric activity--age relation we have used throughout 
            this paper (solid line), 
            the constant-sfr calibration proposed by Soderblom et al. (dotted line) and the constant-sfr calibration 
            from our sample (dashed line). The data and symbols are the same from Soderblom et al. (\cite{soder91}): 
            $\odot$, the sun; H, Hyades; U, Ursa Major Group; open diamonds, field F dwarfs; filled diamonds, 
            binaries.}
      \label{calibs}
      \end{figure}

Presently, there is good indication that the chromospheric activity of a star is linked with its rotation, 
and that the rotation rate decreases slowly with time. However, it is unknown how exactly the chromospheric activity 
is set and how it develops during the stellar lifetime. The data show that {\it there is} a chromospheric 
activity--age relation, but the scatter is such that it is not presently possible to know whether the chromospheric 
activity decreases steadily with time, or there are plateaux around some `preferred' activity levels. There is 
a possibility that the clumps we are seeing in the chromospheric age distribution (which are further identified as 
bursts) are artifacts produced by a monotonic chromospheric activity--age relation.

To keep the constancy of the SFH, Soderblom et al. (\cite{soder91}) proposed an alternative chromospheric 
activity--age relation that is highly non-monotonic. We have checked this constant-sfr calibration with our 
sample, but the result is not a constant sfr. This is expected, since there are many differences in the 
chromospheric samples used by Soderblom (\cite{soder}) and Soderblom et al. (\cite{soder91}) and the one we have used 
(see our Figure 11 in Paper I). We have calculated a new constant-sfr calibration, in the way outlined by 
Soderblom et al. (\cite{soder91}). We have used 328 stars from our sample (just the stars with solar metallicity, to avoid 
the metallicity dependence of $\log R'_{\rm HK}$), with weights given by the volume correction (to account 
for the completeness of the sample) and using the scale height correction factors to take into account the disk 
heating.

Figure \ref{calibs} compares the chromospheric activity--age relation we have used (solid line) 
with the constant-sfr calibration proposed by Soderblom et al. (dotted line) and the constant-sfr 
calibration from our sample (dashed line). The data and symbols are the same from Soderblom et al. (\cite{soder91}). 
Both constant sfr calibrations agree reasonably well for the active stars, but deviate somewhat for the 
inactive stars. This is caused by the fact that to be consistent with a constant sfr, the calibration must 
account for the increase in the relative proportions of inactive to active stars, especially around 
$\log R'_{\rm HK} = -4.90$, 
after the survey of HSDB. Note that, our constant-sfr chromospheric activity--age relation is still barely consistent 
with the data and cannot be ruled out. There are few data for stars older than the Sun in the plot, and it is not possible 
to know whether the plateau for $\log R'_{\rm HK}< -5.0$ in this calibration is real or not.

We acknowledge that, {\it given no other information}, it is a subjective matter whether to prefer a complex star 
formation history or a complex activity-age relation. Nevertheless, 
there are numerous independent lines of evidence that also point to a bursty star formation history; the most 
recent and convincing is the paper by Hernandez, Valls-Gabaud \& Gilmore (\cite{valls}). They use a totally different 
technique (colour--magnitude diagram inversion) and find clear signs of irregularity in the star formation. 
In section 6, we listed 
several other works that indicate a non-constant star formation history, and the majority of them use different 
assumptions and samples. Strongly discontinuous star formation histories are also found for some galaxies in 
the Local Group (see O'Connell \cite{oconnell}), in spite of the initial expectations during the early studies of 
galactic evolution that these galaxies should have had smooth star formation histories.

For all these methods to give the same sort of result, {\it all} the different kinds of calibrations would have to 
contain complex structure. It is simpler to infer that the star formation history is the one that actually has 
a complex structure. We think that when several independent methods all give a similarly bursty star formation 
history (although with different age calibrations, so they do not match exactly), our conclusion is supported over 
the irregular activity-age but constant star formation rate solution.

\section{Conclusions}
     
     A sample composed of 552 stars with chromospheric ages and photometric 
     metallicities was used in the derivation of the star formation history 
     in the solar neighbourhood. Our main conclusions can be summarized as follow:

    \begin{enumerate}
    \item Evidence for at least three epochs of enhanced star formation in the Galaxy were 
     found, at 0-1, 2-5 and 7-9 Gyr ago. These `bursts' are similar to the ones previously 
     found by a number of other studies. 
    \item We have tested the correlation between the SFH and the metal-enrichment rate, given 
     by our AMR derived in Paper I. We have found no correlation between these parameters, 
     although the use of Fe as a metallicity indicator, and the magnitude of the errors 
     in both functions can still hinder the test.
    \item We examined in some detail the possibility 
     that the Galactic bursts are coeval with features in the star formation history of  
     the Magellanic Clouds and close encounters between them and our Galaxy. While the 
     comparison is still uncertain, it points to interesting coincidences that merit 
     further investigation. 
    \item A number of simulations was done to measure the probability for the features 
     found to be consistent with a constant SFH, in face of the age errors that smear out 
     the original features. This probability is shown to decrease for 
     the younger features (being nearly 0\% for the quiescence in the SFH between 1-2 Gyr), 
     such that we cannot give a strong assertion about the burst at 7-9 Gyr. 
     On the other hand, the simulations allow us to conclude, with more than 98\% of confidence, 
     that the SFH of our Galaxy was not constant. 
    \end{enumerate}

      There is plenty of room for improvement in the use of chromospheric ages to 
      find evolutionary constraints. For instance, a reconsideration of the age calibration 
      and a better estimate of the metallicity corrections could diminish  
      substantially the age errors, which would not only improve the age determination 
      but also give more confidence in the older features in the recovered SFH.

\begin{acknowledgements}
      We thank Johan Holmberg for kindly making his data on the scale heights available to us 
      before publication, and Eric Bell for a critical reading of the manuscript, and for giving important 
      suggestions with respect to the presentation of the paper. The referee, 
      Dr. David Soderblom, has raised several points, which contributed to 
      improve the paper. 
      We have made extensive use of the SIMBAD database, operated at CDS, 
      Strasbourg, France. This work was supported by FAPESP and CNPq to WJM and HJR-P,
      NASA Grant NAG 5-3107 to JMS, and the Finnish Academy to CF. 
\end{acknowledgements}

\end{document}